\documentclass[12pt]{article}
\usepackage[margin=1in]{geometry}
\usepackage{natbib}
\usepackage{array,graphicx}
\usepackage[figuresleft]{rotating}
\usepackage{booktabs}
\usepackage{placeins}
\usepackage{amsmath}
\usepackage{amssymb}
\usepackage{amsfonts}
\usepackage{multirow}
\usepackage{amsthm}
\usepackage{bm}
\usepackage{times}
\usepackage{mathabx}
\usepackage{algorithm}
\usepackage{algpseudocode}
\usepackage{enumitem}
\usepackage{setspace}
\usepackage{xurl}
\usepackage[hypertexnames=false]{hyperref}
\hypersetup{colorlinks=true, linkcolor=blue, citecolor=blue, urlcolor=blue}

\newtheorem{theorem}{Theorem}
\newtheorem{lemma}{Lemma}

\newtheorem{proposition}{Proposition}
\theoremstyle{definition}
\newtheorem{definition}{Definition}

\newtheorem{remark}{Remark}

\newcommand{\dM}{\mathcal{D}_{\mathrm{M}}}

\newcounter{Rcount}
\newcommand{\ordR}[1]{%
  \stepcounter{Rcount}%
  \bm{R}^{(\theRcount)}_{n^{#1}}%
}

\renewcommand{\theequation}{\thesection.\arabic{equation}}

\title{Simultaneous Prediction for Latent Variables in Factor Analytic Models}
\author{Zhining Wang\thanks{E-mail: Zhining.Wang@anu.edu.au}, Francis K.C. Hui, Emi Tanaka, A.H. Welsh\\[6pt]
Research School of Finance, Actuarial Studies and Statistics\\
The Australian National University}
\date{}

\begin{document}

\maketitle

\begin{abstract}
Factor analytic models are widely used to analyse multivariate data, producing a set of low-dimensional predicted factor scores (also referred to as latent variables) that can be interpreted or compared across observational units. However, standard pointwise uncertainty measures do not provide simultaneous coverage when factor scores are examined or compared jointly across multiple units. To address this problem, we propose simultaneous prediction regions (SPRs) for latent variables and more general mixed parameter targets in factor analytic models. The SPRs are based on Mahalanobis-type prediction errors, and calibrated through a parametric bootstrap approximation to the distribution of the maximum statistic over all prediction targets. The formulation allows us to accommodate multidimensional targets, such as models with more than one latent factor. 
We establish the asymptotic validity of the proposed bootstrap SPRs procedure under a setting where the number of prediction targets grows faster than the
number of units. Simulation studies demonstrate that the proposed bootstrap SPRs provide more reliable simultaneous coverage across a range of sample sizes and factor-signal strengths compared with simpler Monte-Carlo and Bonferroni alternatives, while an application to ratings of coffee bean quality 
illustrate how SPRs facilitate joint comparisons of country level latent coffee quality ratings.

\vspace{9pt}
\end{abstract}

\section{Introduction}
\label{sec:introduction}
Factor analytic (FA) models, also commonly referred as latent variable models, are widely used for analysing multivariate data.
FA models represent the dependence among a large number of observed variables via a set of low-dimensional factor scores or latent variables \citep{skrondal2004generalized,bartholomew2011latent}. 
After model fitting, the predicted factor scores locate each observational unit in the latent space, while the estimated loading matrix describes the relationship between the observed and latent variables.

In many applications of FA models, the latent variables are not just intermediate quantities in a dimension reduction process, but also serve as important quantities for interpretation and comparison. For example, in psychometrics the factor scores are commonly used to represent cognitive, or personality traits, and may be subsequently compared across individuals or used in downstream statistical analyses \citep{distefano2009understanding, agelink2020factor}. 
In plant breeding trials, FA models are used to model genotype and environment interaction \citep{smith2015factor}, while in community ecology, FA models are readily used to perform model-based ordination with the resulting latent variables (representing sites on a low-dimensional species relative abundance space) examined to look for outlying sites or for assessing effects of intervention/climate effects \citep[e.g.,][]{van2021model}.

Since factor scores are unobserved random quantities, inference on them naturally becomes a prediction problem, analogous to prediction inference for random effects in mixed effects models. On the other hand, much of the existing methodological and applied research on FA models reports the predicted factor score of each observational unit and quantifies its associated uncertainty in a pointwise manner. For instance, prediction intervals for each factor score are constructed separately, without addressing the joint uncertainty that arises when scores are examined simultaneously across multiple observational units \citep[see,][among many others]{grice2001computing,hoshino2008standard,distefano2009understanding,van2021model,korhonen2023fast}. This pointwise perspective is also reflected in commonly used software. For instance, the \texttt{lavaan} package in \texttt{R} provides standard errors for individual predicted factor score, 
from which pointwise intervals can be constructed \citep{rosseel2012lavaan}; a similar approach is used for linear and generalized linear mixed models in the \texttt{lme4} package \citep{bates2015fitting}.
This distinction is important because in many applications of FA models, the interest is not just in the factor score of a single observational unit, but in scores across multiple (if not all) observational units or more generally across a collection of mixed parameter targets, e.g., pairwise differences of the factor scores; note in such cases it is possible for the number of targets to exceed the number of units.
While a pointwise prediction interval, or more generally a prediction region if we are considering a multidimensional mixed parameter target, may attain the nominal coverage level for each target individually, this does not imply simultaneous coverage over the full collection of targets. Indeed, the joint coverage probability can be substantially smaller than the nominal level when the number of targets is large. 

To address the above issues, simultaneous inference methods can be used to perform inference jointly on a collection of targets. The vast majority of research into simultaneous inference has primarily focused on fixed effects parameters in (high-dimensional) regression models or some variation thereof; see \citet{miller2012simultaneous,benjamini2010simultaneous} for general introductions on this topic, along with \citet{zrnic2024locally,du2025simultaneous} 
for examples of recent research in this area, noting that our focus is connected to but distinct from that of the multiple/sequential testing field 
\citep[see][and references therein]{cui2021handbook}. Sporadic developments have also appeared specifically in the FA model literature: \citet{dickhaus2019simultaneous} developed simultaneous testing procedures for dynamic factor models based on multiplicity-adjusted chi-square approximations
while \citet{wang2020simultaneous} considered simultaneous inference for factor loadings in high-dimensional approximate factor models 
These approaches, however, are concerned with simultaneous hypothesis testing of model parameters rather than prediction of latent factor scores. We also note the related work in  simultaneous inference for fixed effects in linear mixed models, e.g., \citet{ritz2017simultaneous} for multilevel models, and \citet{jensen2018comparison} for comparisons of joint and marginal modelling approaches.

More recently, \cite{reluga2023simultaneous,reluga2024bootstrap} and \cite{reluga2023simultaneousGLMM} propose bootstrap-based simultaneous prediction of mixed parameter targets in linear and generalised linear mixed models, respectively. However, their principal prediction targets are \emph{scalar} mixed parameter targets. To the best of our knowledge, simultaneous prediction for multidimensional targets, notably latent variables in FA models, remains underdeveloped.

In this paper, we develop bootstrap-based simultaneous prediction regions (SPRs) for FA models. We formulate the inferential target as a collection of general multidimensional mixed parameters, which includes the special and common cases of predicted factor scores (both one- or multi-dimensional), pairwise differences of factor scores, and point predictions involving linear combinations of fixed effects and latent scores. For each target, we quantify uncertainty through a prediction mean squared error covariance matrix, and use bootstrapping to approximate the distribution of a maximum Mahalanobis-type statistic over all targets, yielding a common critical value that provides prediction regions with asymptotically correct simultaneous coverage.
Under a general setting where the number of targets can grow faster than the number of 
observational units, we establish large sample guarantees for the proposed bootstrap SPRs. Simulations demonstrate the strong finite-sample performance of bootstrap SPRs in comparison with a simpler Bonferroni correction and a Monte-Carlo sampling approach. 
Simulations also verify that the standard pointwise prediction region technique commonly seen in the literature and software consistently and substantially undercovers when it comes to simultaneous coverage. We illustrate an application of our bootstrap SPRs to perform simultaneous prediction on a dataset containing ratings of coffee bean quality, where country-level sensory ratings are represented through one- and two-factor FA models. The resulting simultaneous prediction regions/intervals show how uncertainty can change the interpretation of apparent differences among countries, and facilitate valid joint inference among multidimensional latent scores.

The remainder of the paper is organised as follows. Section~\ref{sec:review-fa} reviews the FA model, relevant inferential quantities, and defines the general mixed parameter targets. Section~\ref{sec:bootstrap} formulates the proposed bootstrap SPR approach and establishes its theoretical properties, while Section~\ref{sec:alternative} discusses two alternative methods based on a Monte-Carlo method and a Bonferroni correction. Section~\ref{sec:simulation} presents results from a simulation study, while Section~\ref{sec:applications} illustrates the proposed method on the coffee bean dataset. Finally, Section~\ref{sec:discussion} offers some concluding remarks, with computational details available at the end. \texttt{R} code for fitting the FA model and implementing the bootstrap SPR approach is available at \url{https://github.com/ZhiningW/repo_simultaneous_prediction_FA}. 

\section{FA model and prediction targets}
\label{sec:review-fa}
Consider a set of $n$ independent observational units $\{(\bm{x}^\top_i, \bm{y}^\top_i)^\top; i = 1,\ldots,n\}$, where $\bm{y}_i = (y_{i1},\ldots,y_{ip})^\top$ denotes a $p$-vector of observed variables and $\bm{x}_i = (x_{i1},\ldots,x_{iq})^\top$ a $q$-vector of covariates. Both $p$ and $q$ are assumed to be fixed throughout this paper. We focus on the general FA model
\begin{align}
    y_{ij} = \bm{x}_i^\top \bm{\beta}_j + \bm{f}_i^\top \bm{\lambda}_j + \epsilon_{ij},
    \label{eq:classic-FR}
\end{align}
where $\bm{\beta}_j$ denotes the coefficient vector for the $j$th response, $\bm{f}_i$ is an $m$-dimensional latent variable for the $i$th observational unit, with $m \ll p$, and $\bm{\lambda}_j$ denotes the corresponding loadings for the $j$th response. 
Finally, $\epsilon_{ij}$ denotes independent error terms that are independent of the latent variables and independently normally distributed, i.e., $\epsilon_{ij} \sim \mathcal{N}(0, \psi_{jj})$, where $\psi_{jj}$ are response-specific error variances (also known as uniquenesses). As is common in many FA model formulations \citep{skrondal2004generalized}, we assume the latent variables follow a standard normal distribution, i.e., $f_{ik} \sim \mathcal{N}(0,1)$ for all $k = 1,\dots,m$.  
Model \eqref{eq:classic-FR} and the bootstrap SPR approach proposed below can be straightforwardly adapted to incorporate other regression terms such as covariates/coefficients that are common to all responses $\bm{x}_i^\top\bm{\beta}$, and also to response-specific covariates $\bm{x}_j$ \citep[e.g.,][]{sammel2002effects}.

From the FA model in \eqref{eq:classic-FR}, we can show that $\bm{V} = \mathrm{Cov}(\bm{y}_i) = \bm{\Lambda}\bm{\Lambda}^\top + \bm{\Psi}$, where $\bm{\Lambda} = \left[\bm{\lambda}_1 \; \bm{\lambda}_2 \; \cdots \; \bm{\lambda}_p\right]^\top$ denotes the $p \times m$ loading matrix, and $\bm{\Psi} = \mathrm{Diag}(\psi_{11}, \psi_{22}, \dots, \psi_{pp})$ is the diagonal matrix of uniquenesses. That is, the FA model induces a reduced-rank form for the covariance matrix between responses, and thus acts as a form of model-based dimension reduction. 
On the other hand,
without imposing additional constraints \eqref{eq:classic-FR} is not identifiable since any right multiplication of $\bm{\Lambda}$ by an $m \times m$ orthogonal rotation
will yield the same marginal likelihood function. 
In this paper, we adopt a standard corner constraint \citep{skrondal2004generalized,korhonen2023fast}.

To facilitate our developments below, we express the FA model in a fully vectorised form as
\begin{align}
\bm{y} = (\bm{I}_p \otimes \bm{X})\bm{\beta} + (\bm{\Lambda} \otimes \bm{I}_n)\bm{f} + \bm{\epsilon},
\label{eq:vectorised-FR}
\end{align}
where $\bm{y} = (y_{11},\ldots,y_{n1},y_{12},\ldots,y_{np})^\top$ is the full $np$-vector of responses, $\bm{X}$ is the $n \times q$ matrix of covariates formed by stacking the $\bm{x}_i$'s as row vectors, $\bm{\beta} = (\bm{\beta}_1^\top,\ldots,\bm{\beta}_p^\top)^\top$, $\bm{f} = (f_{11},\ldots,f_{n1},f_{12},\ldots,f_{nm})^\top,$ and
$\bm{\epsilon} = (\epsilon_{11},\ldots,\epsilon_{n1},\epsilon_{12}, \ldots,\epsilon_{np})^\top$.
Under this formulation, it follows that $\bm{f} \sim \mathcal{N}_{nm}(\bm{0}_{nm}, \bm{I}_{nm})$ and $\bm{\epsilon} \sim \mathcal{N}_{np}(\bm{0}_{np},\, \bm{\Psi} \otimes \bm{I}_n)$, where $\bm{0}_c$ denotes the $c$-dimensional zero vector and $\bm{I}_c$ denotes the $c \times c$ identity matrix.

\subsection{Pointwise predictive regions} \label{subsec:ppr}
Turning to prediction, we define a $d$-dimensional general mixed parameter 
\begin{align}
\bm{\mu}_l = \bm{C}_l^\top \bm{\beta} + \bm{K}_l^\top \bm{f}, \qquad l = 1,\dots,L,
\label{eq:def-multid-gmp}
\end{align}
where $\bm{C}_l \in \mathbb{R}^{pq \times d}$ and $\bm{K}_l \in \mathbb{R}^{nm \times d}$ are user-specified, fixed design matrices. Definition \eqref{eq:def-multid-gmp} can be used to represent a broad range of prediction targets, with some notable examples being: 
1) the factor scores themselves, by setting $L = n, d = m$, $\bm{C}_l = \bm{0}_{pq \times d}$ for all $l$, and $\bm{K}_l$ to a sparse matrix of zeros and ones such that $\bm{K}_l^\top\bm{f} = (f_{i1},\ldots,f_{im})^\top$;
2) pairwise differences between factor scores, by setting $L= n(n-1)/2$, $\bm{C}_l = \bm{0}_{pq \times d}$ but $\bm{K}_l$ to a sparse matrix comprising elements $\{-1,0,1\}$ such that $\bm{K}_l^\top\bm{f} = (f_{i1}-f_{i'1},\ldots,f_{im} - f_{i'm})^\top$ for $i \ne i'$, and;
3) general predictions for individual observations.

Under the FA model \eqref{eq:classic-FR}--\eqref{eq:vectorised-FR}, it is straightforward to show that the maximum likelihood estimator (MLE) of $\bm{\beta}$ is given by the ordinary least squares estimator $\widehat{\bm{\beta}} = \left( \bm{I}_p \otimes (\bm{X}^\top \bm{X})^{-1} \bm{X}^\top \right)\bm{y}$. Following standard linear mixed model theory for random (mixed) effect prediction then \citep{searle2009variance}, given the variance components $\bm{\theta} = ( \mathrm{vech}(\bm{\Lambda})^\top, \psi_{11}, \ldots, \psi_{pp} )^\top$, where $\mathrm{vech}(\bm{\Lambda})$ extracts the lower triangular elements of the loading matrix and noting the diagonal elements are constrained to be positive, the best linear unbiased predictor (BLUP) 
of the factor scores is given by
\begin{align*}
\widetilde{\bm{f}} = \widetilde{\bm{f}}(\bm{\theta}) = 
\left( \bm{\Lambda}^\top \bm{V}^{-1} \otimes \bm{I}_n \right)\left( \bm{y} - (\bm{I}_p \otimes \bm{X})\widehat{\bm{\beta}} \right).
\end{align*}
The BLUP of the general mixed parameter then follows as  $\widetilde{\bm \mu}_l = \bm C_l^\top \widehat{\bm \beta} + \bm K_l^\top \widetilde{\bm f}$.
In practice, the variance components need to be estimated e.g., we used MLE in the simulation studies and application in Sections \ref{sec:simulation} and \ref{sec:applications}, although restricted maximum likelihood estimation (REML) is a popular alternative. 
Regardless, on replacing $\bm{\theta}$ by its estimate $\widehat{\bm{\theta}}$, we obtain the empirical BLUP (EBLUP) of the general mixed parameter as $\widehat{\bm{\mu}}_l = \widetilde{\bm{\mu}}_l(\widehat{\bm{\theta}}) = \bm{C}_l^\top \widehat{\bm{\beta}} + \bm{K}_l^\top \widehat{\bm{f}}$, where $\widehat{\bm f} = \widetilde{\bm f}(\widehat{\bm\theta})$ denotes the EBLUP of $\bm f$. 
 
Turning to uncertainty quantification for general mixed parameters, in linear mixed models, a common approach is to estimate the prediction mean squared error (PMSE) and construct regions using this covariance matrix. 
Specifically, given the variance component $\bm{\theta}$, straightforward algebraic calculations show that the PMSE of the BLUP $\widetilde{\bm{\mu}}_l$ 
is given by
\begin{align}
\bm{\Sigma}(\widetilde{\bm{\mu}}_l; \bm{\theta}) &= \mathbb{E}\{(\widetilde{\bm{\mu}}_l - \bm{\mu}_{0,l})(\widetilde{\bm{\mu}}_l - \bm{\mu}_{0,l})^\top\} \nonumber \\
&= \bm{K}_l^\top\!\left(\bm{I}_{nm} - (\bm{\Lambda}^\top \bm{V}^{-1}\bm{\Lambda}) \otimes \bm{I}_n\right)\!\bm{K}_l + \bm D_l^\top \bm{Q} \bm D_l \nonumber \\
&=: \bm{G}_{1,l}(\bm{\theta}) + \bm{G}_{2,l}(\bm{\theta}), \label{eq:g1g2}
\end{align}
where $\bm D_l = \bm{C}_l - (\bm{I}_p \otimes \bm{X}^\top)\bm{O}_l$, $\bm{O}_l = (\bm{V}^{-1}\bm{\Lambda} \otimes \bm{I}_n)\bm{K}_l$, and  $\bm{Q} = \bm{V} \otimes (\bm{X}^\top \bm{X})^{-1}$. Meanwhile, we use $\bm{\mu}_{0,l} = \bm C_l^\top \bm\beta_0 + \bm K_l^\top \bm f_0$ to denote the underlying true general mixed parameter.
The term $\bm{G}_{1,l}(\bm{\theta})$ represents the uncertainty in predicting $\bm{f}$ by its BLUP when $\bm{\beta}$ is known, while $\bm{G}_{2,l}(\bm{\theta})$ accounts for the additional uncertainty in estimating $\bm{\beta}$ by its estimator $\widehat{\bm{\beta}}$. 
Turning to the EBLUP,
we have the further decomposition $\bm{\Sigma}(\widehat{\bm{\mu}}_l; \bm \theta) = \mathbb{E}\{(\widehat{\bm{\mu}}_l - \widetilde{\bm{\mu}}_l)(\widehat{\bm{\mu}}_l - \widetilde{\bm{\mu}}_l)^\top\} + \bm{\Sigma}(\widetilde{\bm{\mu}}_l; \bm{\theta}) - 2\,\mathbb{E}\left\{(\widehat{\bm{\mu}}_l - \widetilde{\bm{\mu}}_l)(\bm{\mu}_l - \widetilde{\bm{\mu}}_l)^\top \right\}$, where the first term captures the additional variability due to estimation of the variance components and is generally analytically intractable, and the third cross term is known to be of a smaller order under the regularity conditions considered below \citep[see][]{searle2009variance,reluga2023simultaneous}.
With this in mind, in this paper we propose to approximate the PMSE of $\widehat{\bm\mu}_l$ by the PMSE of $\widetilde{\bm\mu}_l$ \eqref{eq:g1g2} evaluated at $\widehat{\bm\theta}$, that is, 
$\widehat{\bm{\Sigma}}(\widehat{\bm{\mu}}_l) \approx \bm{\Sigma}(\widetilde{\bm{\mu}}_l; \widehat{\bm{\theta}})$. 

We now formally define a pointwise prediction region (PPR) for the
multidimensional general mixed parameter as follows \citep[note][refer to these as clusterwise prediction intervals in the context of linear mixed models]{reluga2023simultaneous}.

\begin{definition} \label{def:ppr}
Let $\{\bm{\mu}_{0,l}; l = 1,\ldots,L\}$ denote the collection of true $d$-dimensional multidimensional general mixed parameters. Then for a nominal level $1-\alpha$, the PPR is defined as the Cartesian product region
$\mathcal{P}_{1-\alpha} = \bigtimes_{l=1}^L \mathcal{P}_{1-\alpha,l}$, where each region $\mathcal{P}_{1-\alpha,l}\subset \mathbb{R}^d$ is calibrated such that $\mathbb{P}\left(\bm{\mu}_{0,l} \in \mathcal{P}_{1-\alpha,l} \right) \rightarrow 1-\alpha$ as $n \to \infty$ for all $l = 1,\ldots,L$.
\end{definition}

To construct PPRs in Definition \ref{def:ppr}, we can use the fact that, for each fixed target $l$, the studentised statistic $\bm{T}_{l,n} = \widehat{\bm{\Sigma}}(\widehat {\bm\mu}_l)^{-1/2} \{\widehat{\bm{\mu}}_l - \bm{\mu}_{0,l}\}$ converges to a $d$-dimensional multivariate standard normal distribution under regularity conditions (see Section~\ref{subsec:theory}); see SM~\ref{SM-sec:prop1} for a proof of this. This suggests that we can use 
a Mahalanobis-type distance $\dM^2(\widehat{\bm\mu}_l,\bm\mu_{0,l};\widehat{\bm\Sigma}(\widehat{\bm\mu}_l))$, where $\dM(\bm v_1,\bm v_2;\bm\Sigma)= \left\{(\bm v_1-\bm v_2)^\top\bm\Sigma^{-1}(\bm v_1-\bm v_2) \right\}^{1/2}$ with $\bm v_1,\bm v_2\in\mathbb R^c$ and a positive definite matrix $\bm\Sigma\in\mathbb R^{c\times c}$, to construct the PPR.
Moreover, the above asymptotic normality implies $\dM^2(\widehat{\bm\mu}_l,\bm\mu_{0,l};\widehat{\bm\Sigma}(\widehat{\bm\mu}_l)) \xrightarrow{d} \chi_d^2$ for each $l=1, \ldots, L$, where $\chi_d^2$ denotes the chi-square distribution with $d$ degrees of freedom (DoF); see also \cite{ito2021corrected} for a similar application in the context of the multivariate Fay-Herriot models.
We thus conclude that the PPR for a single true general mixed parameter $\bm{\mu}_{0,l}$ is given by the ellipsoidal region 
\begin{align*}
\widehat{\mathcal{P}}_{1-\alpha,l} = \left\{\bm{x} \in \mathbb{R}^d: \dM \left(\widehat{\bm{\mu}}_l, \bm{x};\widehat{\bm{\Sigma}}\left(\widehat{\bm{\mu}}_l\right) \right) \leq c^{\mathrm{PPR}}_{1-\alpha}\right\},
\end{align*}
where $c^{\mathrm{PPR}}_{1-\alpha}= \sqrt{F^{-1}_{\chi^2_d}\!\left(1 - \alpha \right)}$ denotes the so-called pointwise critical value and is defined as the square root of the $(1-\alpha)$th quantile of a 
chi-square distribution with DoF $d$. The PPR for the collection in Definition \ref{def:ppr} follows immediately as $\widehat{\mathcal{P}}_{1-\alpha} = \bigtimes_{l=1}^L \widehat{\mathcal{P}}_{1-\alpha,l}$. 

\section{Bootstrap simultaneous prediction regions}
\label{sec:bootstrap}

While Definition \ref{def:ppr} and the estimated PPRs offer a standard approach for performing inference on each individual mixed parameter, when the collection $\{\bm{\mu}_{0,l}; l = 1, \ldots, L\}$ is considered jointly as is often the case in FA models (see Section \ref{sec:introduction}) then the pointwise coverage is no longer adequate. That is, while each $\mathcal P_{1-\alpha,l}$ has nominal coverage $1-\alpha$ for a fixed $l$, the probability that all $L$ regions simultaneously cover their corresponding targets is smaller than $1-\alpha$, with the discrepancy increasing as $L$ increases. To address this problem, and motivated by the work of \citet{reluga2023simultaneous}, we propose simultaneous prediction regions (SPRs) for FA models defined as follows.

\begin{definition} \label{def:spr}
Let $\{\bm{\mu}_{0,l}; l = 1,\ldots,L\}$ denote the collection of true $d$-dimensional multidimensional general mixed parameters. Then for a nominal level $1-\alpha$, the SPR is defined as the Cartesian product region $\mathcal S_{1-\alpha} = \bigtimes_{l=1}^L\mathcal S_{1-\alpha,l}$, where each region $\mathcal S_{1-\alpha,l}\subset\mathbb R^d$ is calibrated such that the collection satisfies $\mathbb P(\bigcap_{l=1}^L\{\bm\mu_{0,l}\in\mathcal S_{1-\alpha,l}\}) \to 1-\alpha$ as $n\to \infty$.
\end{definition}

SPR calibration therefore accounts for the multiplicity induced by the full collection of mixed parameters. Note we retain the ellipsoidal form of the PPRs in Definition~\ref{def:spr}, but use a common critical value calibrated instead based on the following maximum statistic: for a candidate value $c$, define
$
\mathcal S_{1-\alpha,l}(c)=\{\bm x\in\mathbb R^d:\dM(\widehat{\bm\mu}_l,\bm x;\widehat{\bm\Sigma}(\widehat{\bm\mu}_l))\leq c \} \text{ for } l=1,\dots, L
$. 
Then using Definition \ref{def:spr}, we can write
\begin{align*}
\mathbb P\left(\bigcap_{l=1}^L\{\bm\mu_{0,l}\in\mathcal S_{1-\alpha,l}(c)\}\right)
&= \mathbb P\left\{\dM\left(\widehat{\bm\mu}_l,\bm\mu_{0,l};\widehat{\bm\Sigma}(\widehat{\bm\mu}_l)\right)\leq c,\; l=1,\ldots,L\right\}\\
&=\mathbb P\left\{\max_{1\leq l\leq L}\dM\left(\widehat{\bm\mu}_l,\bm\mu_{0,l};\widehat{\bm\Sigma}(\widehat{\bm\mu}_l)\right)\leq c\right\}.
\end{align*}
It follows that we can use the maximum Mahalanobis-type statistic
$\tau=\max_{1\leq l\leq L}\tau_l$ with 
$\tau_l=\dM(\widehat{\bm\mu}_l,\bm\mu_{0,l};\widehat{\bm\Sigma}(\widehat{\bm\mu}_l))$, and construct a new critical value as
$c^{\mathrm{SPR}}_{1-\alpha}=\inf\{t\in\mathbb R:\mathbb P(\tau\leq t)\geq 1-\alpha\}$.

In contrast to the PPR case in Section \ref{subsec:ppr}, the critical value $c^{\mathrm{SPR}}_{1-\alpha}$ does not possess a closed-form expression, since the statistics $\{\tau_l; l = 1, \ldots, L\}$ are dependent and are all functions of the same estimated regression coefficients, factor scores, and variance components. 
With this in mind, we propose to use the parametric bootstrap to approximate the joint distribution of $\{\tau_l; l = 1, \ldots, L\}$ and hence its maximum $\tau$.
In detail, under the parametric bootstrap, the fitted FA model is treated as the data-generating mechanism, with estimates given by $(\widehat{\bm\theta}^\top,\widehat{\bm\beta}^\top)^\top$. Then, in the $b$th bootstrap sample, we generate new factor scores $\bm f^{*(b)} \sim \mathcal N_{nm}(\bm0_{nm},\bm I_{nm})$ and responses $\bm y^{*(b)}=(\bm I_p\otimes\bm X)\widehat{\bm\beta} + (\widehat{\bm\Lambda}\otimes\bm I_n)\bm f^{*(b)} + \bm\epsilon^{*(b)}$ with $\bm\epsilon^{*(b)} \sim \mathcal N_{np}(\bm0_{np},\widehat{\bm\Psi}\otimes\bm I_n)$, and set $\bm\mu_l^{*(b)}=\bm C_l^\top\widehat{\bm\beta}+\bm K_l^\top\bm f^{*(b)}$ as the bootstrap version of the true general mixed parameter $\bm \mu_{0,l}$, for $l = 1, \ldots, L$.
Next, we refit the FA model \eqref{eq:vectorised-FR} to the bootstrapped responses to obtain bootstrap estimates $(\widehat{\bm\theta}^{*(b)\top}$, $\widehat{\bm\beta}^{*(b)\top})^\top$, and subsequently the bootstrap EBLUP of the general mixed parameter $\widehat{\bm\mu}_l^{*(b)} = \bm C_l^\top\widehat{\bm\beta}^{*(b)}+\bm K_l^\top\widehat{\bm f}^{*(b)}$ and associated PMSE $\widehat{\bm\Sigma}^{*(b)}(\widehat{\bm\mu}_l^{*(b)})$, for $l=1,\ldots,L$. Afterwards, we compute the bootstrap analogue of the maximum Mahalanobis-type statistic as $\tau^{*(b)}=\max_{1\leq l\leq L}\tau_l^{*(b)}$, where $\tau_l^{*(b)}=\dM (\widehat{\bm\mu}_l^{*(b)},\bm\mu_l^{*(b)};\widehat{\bm\Sigma}^{*(b)}(\widehat{\bm\mu}_l^{*(b)}))$,   
The required bootstrap critical value is then given by 
$\widehat{c}_{1-\alpha}^{\mathrm{SPR,BS}}=\inf\{t\in\mathbb R:\mathbb P^*(\tau^*\leq t)\geq 1-\alpha\}$, and computed as the empirical $(1-\alpha)$-quantile of the bootstrap maximum statistics $\{\tau^{*(b)}; b = 1,\ldots,B\}$. 
Finally, the estimated bootstrap SPR for the collection of general mixed parameters $\mu_l$ in Definition \ref{def:spr} is given by $\widehat{\mathcal S}_{1-\alpha}^{\mathrm{BS}}=\bigtimes_{l=1}^L \widehat{\mathcal S}_{1-\alpha,l}^{\mathrm{BS}}$, where
\begin{align}
    \widehat{\mathcal{S}}_{1-\alpha,l}^{\mathrm{BS}}=\left\{\bm x\in\mathbb R^d:\dM\left(\widehat{\bm\mu}_l,\bm x;\widehat{\bm\Sigma}(\widehat{\bm\mu}_l)\right)\leq \widehat{c}_{1-\alpha}^{\mathrm{SPR,BS}}\right\}.
    \label{eq:spr-bs}
\end{align}
A formal algorithm is given in SM~\ref{SM-sec:detailed-algorithm}, noting that prior to computing $\tau^{*(b)}$ in each bootstrapped dataset, some additional processing may be required to handle finite sample misalignment in rotations and ordering; see SM~\ref{SM-subsec:align-bootstrap} for further details of this alignment.

When using the bootstrap procedure in practice, we follow the recommendation of \cite{reluga2023simultaneous,reluga2023simultaneousGLMM} and approximate the PMSE covariance matrix in \eqref{eq:g1g2} only by its leading component, $\bm G_{1,l}(\widehat{\bm\theta})$. Although this choice may lead to mild overcoverage in small samples (see Section~\ref{sec:simulation}) it is theoretically justified by the fact that $\bm G_{2,l}(\widehat{\bm\theta})$ is of smaller order under general regularity conditions; see the proof of Proposition~\ref{SM-prop:normal-approx} for details. More importantly, empirically like \citet{reluga2023simultaneous}, we find that $\bm G_{1,l}(\widehat{\bm\theta})$ tends to produce more stable bootstrap critical values. By contrast, using the full approximation $\bm G_{1,l}(\widehat{\bm\theta})+\bm G_{2,l}(\widehat{\bm\theta})$ often leads to undercoverage when constructing bootstrap SPRs as sample size increases; see Section~\ref{sec:discussion} for further discussion of this behaviour.

\subsection{Asymptotic results}
\label{subsec:theory}

In this section, we establish that the bootstrap critical value estimate $\widehat{c}^{\mathrm{SPR,BS}}_{1-\alpha}$ asymptotically approximates the underlying true critical value $c^{\mathrm{SPR}}$. Let $\zeta_{\min}(\bm{M})$ and $\zeta_{\max}(\bm{M})$ denote the minimum and maximum eigenvalues of a symmetric matrix $\bm{M}$, respectively, and assume that the number of factors $m$ and the dimension of the general mixed parameter $d$ are fixed. Furthermore, assume that the elements in $\bm{\theta}$ corresponding to the diagonal elements of $\bm{\Lambda}$ have been reparameterised, e.g., via a log transformation, such that they are now unconstrained. For ease of notation though, we will not make the reparametrisation explicit in the theoretical developments below. We require the following regularity conditions. 

\begin{enumerate}[label=(A\arabic*), ref=A\arabic*]
    \item The true parameter vector $(\bm{\beta}_0^\top, \bm{\theta}_0^\top)^\top$ is an interior point of the parameter space $\bm{\Theta} \subset \mathbb{R}^{pq + h_\theta}$, where $h_\theta= pm-2^{-1} m(m-1) + p$.
    \label{regcond:true-para}
    \item There exist constants $0< n_0 < \infty$, $0< C_0<\infty$ and $1 <C_1 < \infty$ such that for every $n \ge n_0$ and $i = 1,\ldots,n$, it holds that $\|\bm{x}_i\|_\infty \le C_0$, and $C_1^{-1} \le \zeta_{\min}(n^{-1} \sum_{i=1}^n \bm{x}_i\bm{x}_i^\top) \le \zeta_{\max}(n^{-1}\sum_{i=1}^n \bm{x}_i\bm{x}_i^\top) \le C_1$. \label{regcond:X}
    \item Write $\bm{K}_l^\top = \left[\bm{K}_{l1} \; \bm{K}_{l2} \; \cdots \; \bm{K}_{ln} \right]$ where $\bm{K}_{l1}$ is a $d \times m$ matrix, and define
    $\mathcal{S}_l = \left\{i: \bm{K}_{li} \neq \bm{0}_{d\times m} \right\}$. Then there exist sufficiently large positive constants $S_K, C_K$ and $C_L$ such that $\sup_{1 \leq l \leq L} |\mathcal{S}_l| \leq S_K$ and $\sup_{1 \leq l \leq L} \sup_{1 \leq i \leq n} \|\bm{K}_{li}\| \leq C_K$. Furthermore, $\max_{1\leq l \leq L}\|\bm{C}_l\| = C_L$. \label{regcond:designed-mat}
    \item There exists an open subset $\omega$ of $\mathbb{R}^{h_\theta}$ containing $\bm\theta_0$ such that, for all $j = 1,\ldots,h_\theta$, it holds that $\sup_{\bm\theta \in \omega} \max_{1\leq l\leq L} \| \partial / \partial\theta_j (\bm I_p\otimes\bm X)^\top\bm O_l(\bm\theta) \| \le C_2,
    $
    for some sufficiently large positive constant $C_2$.
    \label{regcond:D-mat}
    \item  In the same locally identified parameterisation, $\|\widehat{\bm{\theta}} - \bm{\theta}_0\| = O_p(n^{-1/2})$ and $\| \widehat{\bm\theta}^* - \widehat{\bm\theta}\| = O_{p^*}(n^{-1/2})$ in probability under the original sampling law. \label{regcond:MLE}
    \item There exist constants $1 < \kappa_1, \kappa_2 < \infty$ such that $\kappa_1^{-1} \le \zeta_{\min}\{\bm{G}_{1,l}(\bm{\theta}_0)\} \le \zeta_{\max}\{\bm{G}_{1,l}(\bm{\theta}_0)\} \le \kappa_1$, for all $l = 1,2, \dots,L$, and $\kappa_2^{-1} \le \zeta_{\min}(\bm{V}_0) \le \zeta_{\max}(\bm{V}_0) \le \kappa_2$. \label{regcond:PMSE}
\end{enumerate}
Assumption~\ref{regcond:true-para} is a standard regularity condition for FA models \citep{anderson1988asymptotic}.
Assumption~\ref{regcond:X} requires the covariates to be uniformly bounded and the empirical design matrix to remain well-conditioned, and is a common fixed-design condition imposed in asymptotic analyses of mixed models and their predictors \citep[e.g.,][]{chatterjee2008parametric}.
Assumption~\ref{regcond:designed-mat} restricts the complexity of the multidimensional general mixed parameters by requiring each target to depend on only finitely many latent-variable components,
while the contribution from the fixed effects is also uniformly bounded. This is analogous to the bounded-dimensional and bounded-coefficient conditions commonly imposed on mixed parameters in linear mixed model simultaneous prediction in \citet{reluga2023simultaneous}, and is satisfied for most targets of interest including those mentioned above, e.g., for simultaneous prediction of factor scores across observational units, we have $|\mathcal{S}_l| = 1$. 
Assumption~\ref{regcond:D-mat} imposes a local smoothness condition on the prediction operator with respect to the variance components in the FA model;  
similar differentiability conditions are imposed in asymptotic analyses of (E)BLUPs in mixed models \citep[e.g.,][]{chatterjee2008parametric,reluga2023simultaneous}. 
Assumption~\ref{regcond:MLE} follows from standard asymptotic results of maximum likelihood estimators in FA and latent variables models\citep{anderson1988asymptotic,huber2004estimation} under both original and conditional bootstrap laws. Note also the OLS estimator used for $\bm\beta$ coincides with its maximum likelihood estimator and, 
that under Assumption~\ref{regcond:X}, it can be shown that $\|\widehat{\bm\beta}-\bm\beta_0\| = O_p(n^{-1/2})$, with the corresponding conditional bootstrap rate. 
Finally, Assumption~\ref{regcond:PMSE} requires both the PMSE covariance matrix and the true marginal covariance matrix of the responses to be well-conditioned.

\begin{theorem}
\label{thm:bootstrap-consistency}
    Assume Conditions \ref{regcond:true-para}--\ref{regcond:PMSE} hold and the size of the collection of general mixed parameter targets satisfies $\log L = o(\sqrt{n})$. Then as $n \to \infty$, it holds that $\sup_{t\in \mathbb{R}} \left| \mathbb{P}^*(\tau^* \le t) - \mathbb{P}(\tau \le t) \right| \xrightarrow{p} 0$, and hence
\begin{align}
    \mathbb{P}\left( \bigcap_{l=1}^L \left\{\bm{\mu}_{0,l} \in \widehat{\mathcal{S}}_{1-\alpha,l}^{\mathrm{BS}} \right\} \right) \rightarrow 1-\alpha.
\end{align}
\end{theorem}

The proof of the above is given in SM~\ref{SM-sec:proof-thm1}. Theorem~\ref{thm:bootstrap-consistency} formally justifies using  bootstrap critical value $\widehat{c}_{1-\alpha}^{\mathrm{SPR,BS}}$, and hence asymptotically correct simultaneous coverage for a growing collection of latent-factor targets. 
Note although we suggested the usage of leading term of PMSE when doing bootstrap, the theorem technically supports either $\bm G_{1,l}(\widehat{\bm\theta})$ or $\bm G_{1,l}(\widehat{\bm\theta}) + \bm G_{2,l}(\widehat{\bm\theta})$ being used in constructing the original and bootstrap statistics; see the proof of the theorem and Lemma~\ref{SM-lemma:order_of_G1andG2} and Lemma~\ref{SM-lemma:order-reminder} for further details of this.
Also, the growth condition accommodates simultaneous prediction of all individual factor vectors ($L=n$) and all pairwise factor contrasts ($L=n(n-1)/2$), among other possibilities. 

\section{Alternative methods for simultaneous prediction}
\label{sec:alternative}
 
In this section, we discuss two alternative, simpler methods for performing simultaneous prediction in FA models, namely a Monte-Carlo simulation based method and a Bonferroni correction approach. In the context of random intercept linear mixed models,  \citet{reluga2023simultaneous} investigated these two approaches alongside a volume-of-tube method \citep{weyl1939volume, hotelling1939tubes} and a technique based on Beran's method \citep{beran1988balanced}. 
However, through numerical studies they found that the latter two approaches performed substantially worse. As such, we restrict our attention to the Monte-Carlo and Bonferroni correction methods.

\subsection{Monte-Carlo method}
\label{subsec:MC-method}

Let $\bm\xi_{0}^\top = (\bm\beta_{0}^\top, \bm f^\top)$ and $\widetilde{\bm\xi}^\top = ( \widehat{\bm\beta}^\top, \widetilde{\bm f}^\top)$. Then the Monte-Carlo approach to simultaneous prediction is based on the following proposition.

\begin{proposition}
\label{prop:distribution-of-delta}
Given the true variance components $\bm\theta$, the difference $\bm\xi_0 -\widetilde{\bm\xi}$ satisfies 
$$
\bm\xi_{0} - \widetilde{\bm\xi} \sim \mathcal N_{qp+nm}\left( \bm0_{qp+nm}, (\bm A^\top\bm W^{-1}\bm A+\bm P)^{-1}\right),$$
where $\bm A = \left[ \bm I_p \otimes\bm X \;\; \bm \Lambda \otimes \bm I_n\right]$, $\bm W = \bm\Psi \otimes \bm I_n$,
and $\bm P = \mathrm{Bdiag}(\bm0_{qp\times qp}, \bm I_{nm})$ where $\mathrm{Bdiag}(\cdots)$ denotes the block diagonal matrix operator.
\end{proposition}
A proof of the above result based on deriving and then manipulating the mixed model equations \citep{thompson2008estimation} of the vectorised formulation of the FA model in \eqref{eq:vectorised-FR} is provided in SM~\ref{SM-sec:proof-prop2}.
Replacing the unknown variance components in Proposition \ref{prop:distribution-of-delta} by their estimates $\widehat{\bm\theta}$, we obtain the plug-in precision $\widehat{\bm\Omega}_{\xi} = \widehat{\bm A}^\top\widehat{\bm W}^{-1}\widehat{\bm A}+\bm P$, where $\widehat{\bm A}=[\bm I_p\otimes\bm X\;\; \widehat{\bm\Lambda}\otimes\bm I_n]$ and $\widehat{\bm W}=\widehat{\bm\Psi}\otimes\bm I_n$. 
The Monte-Carlo method for simultaneously prediction is then constructed as follows. First, we draw $\bm u^{(s)}\sim\mathcal N_{qp+nm}(\bm0_{qp+nm},\widehat{\bm\Omega}_{\xi}^{-1}), s=1,\ldots,S$ for a large number of Monte-Carlo samples $S$. 
Then, for the $l$th multidimensional general mixed parameter target, we define $\bm H_l = \left[ \bm C_l^\top \;\; \bm K_l^\top\right] \in \mathbb R^{d\times (pq+nm)}$ and compute the corresponding Mahalanobis statistic $\tau_l^{(s)} = \{(\bm u^{(s)})^\top \bm H_l^\top \widehat{\bm\Sigma}(\widehat{\bm\mu}_l)^{-1} \bm H_l \bm u^{(s)} \}^{1/2}$ for $s = 1,\ldots,S$. Afterwards, for each Monte-Carlo sample we compute the maximum statistic $\tau^{(s)}=\max_{1\leq l\leq L}\tau_l^{(s)}$. The Monte-Carlo critical value $\widehat c_{1-\alpha}^{\mathrm{SPR,MC}}$ is then taken as the empirical $(1-\alpha)$-quantile of $\{\tau^{(1)},\ldots,\tau^{(S)}\}$, from which SPRs, $\widehat{\mathcal{S}}_{1-\alpha,l}^{\mathrm{MC}}$ can be constructed in an analogous way to \eqref{eq:spr-bs}; see SM~\ref{SM-sec:detailed-algorithm} for a formal algorithm detailing the Monte-Carlo method for simultaneous prediction.  
For the same reasons as in the bootstrap procedure, we recommend using the leading component $\bm G_{1,l}(\widehat{\bm\theta})$ to approximate the PMSE matrix in the Monte-Carlo approach. 

While the Monte-Carlo SPR is expected to be computationally more efficient than the proposed bootstrap SPR in Section \ref{sec:bootstrap}, since it does not require refitting of the FA model, its performance relies more directly on the quality of the normal approximation from Proposition \ref{prop:distribution-of-delta}, and so its finite-sample performance may deteriorate when the parameters exhibit non-negligible bias or mean squared error e.g., for small $n$; see the simulations in Section \ref{sec:simulation} for evidence of this.

\subsection{Bonferroni correction method}

For a collection of $L$ general mixed parameters, the Bonferroni correction is based on a simple adjustment of the critical value from the PPR.
Specifically, we set $c^{\mathrm{SPR,BO}}_{1-\alpha}=\{F^{-1}_{\chi^2_d}(1-L^{-1}\alpha)\}^{1/2}$, and construct the Bonferroni SPR as $\widehat{\mathcal{S}}_{1-\alpha}^{\mathrm{SPR,BO}}=\bigtimes_{l=1}^L\{\bm x:\dM(\widehat{\bm\mu}_l,\bm x;\widehat{\bm\Sigma}(\widehat{\bm\mu}_l))\leq c^{\mathrm{SPR,BO}}_{1-\alpha}\}$.

The Bonferroni-based critical value is closely related to the Gumbel approximation for the maximum marginal chi-square statistic, and we provide further theoretical explanation of this point in SM~\ref{SM-sec:supp-gumbel-bonferroni}. 
On the other hand, aside from being potentially conservative, the finite-sample accuracy of the Bonferroni SPR depends on accurate marginal approximation of the studentised prediction errors, which in turn relies on reliable estimation of the PMSE matrices. The dependence among the target statistics can further affect the degree of conservativeness of the resulting simultaneous regions.
Indeed, when employing the Bonferroni SPRs we recommend using the \emph{full} PMSE approximation $\bm G_{1,l}(\widehat{\bm\theta})+\bm G_{2,l}(\widehat{\bm\theta})$ as given in \eqref{eq:g1g2}, instead of just the first term as in the bootstrap and Monte-Carlo SPR approaches.
As shown in Section~\ref{sec:simulation}, using only $\bm G_{1,l}(\widehat{\bm\theta})$ with a Bonferroni-based critical value often leads to simultaneous undercoverage.

\section{Simulation study}
\label{sec:simulation}

We performed a numerical study to evaluate the finite-sample performance of the proposed bootstrap SPR approach (Section \ref{sec:bootstrap}), comparing it to the alternative Monte-Carlo and Bonferroni SPRs (Section \ref{sec:alternative}), as well as the PPR method (Section \ref{subsec:ppr}). 
Analogous to \citet{reluga2023simultaneous}, we assessed performance using three metrics: 
let $I$ denote the number of simulated datasets and $\bm{\mu}_{0,l,(i)}$ denote the true value of the $l$th general mixed parameter target in simulated dataset $i = 1,\ldots,I$. Then for the bootstrap SPR defined in \eqref{eq:spr-bs}, we compute its empirical coverage probability as $\mathrm{ECP}^{BS} = I^{-1} \sum_{i=1}^I \mathbb{I} \{\bm{\mu}_{0, l,(i)} \in \mathcal{S}^{BS}_{1-\alpha,l,(i)} \forall l = 1,\ldots,L\}$,
the average volume of the SPR defined as $\mathrm{AV}^{BS} = (IL)^{-1} \sum_{i=1}^I \sum_{l=1}^L \nu^{BS}_{l,(i)}$ where $\nu^{BS}_{l,(i)} = \Gamma(d/2+1)^{-1} \pi^{d/2}  (c^{BS}_{1-\alpha,(i)})^d | \widehat{\bm{\Sigma}}_{l,(i)}|^{1/2}$ 
is the volume of the $d$-dimensional ellipsoid $\mathcal{S}^{BS}_{1-\alpha,l,(i)}$, and the average variance of the volume defined as 
$\mathrm{AVV}^{BS} = \{(I-1)L\}^{-1} \sum_{l=1}^L\sum_{i=1}^I ( \nu^{BS}_{l,(i)} - \bar{\nu}^{BS}_l)^2$ with $\bar{\nu}^{BS}_l = I^{-1} \sum_{i=1}^I \nu^{BS}_{l,(i)}$. 
Analogous values of these three metrics are also computed for the Monte-Carlo and Bonferroni SPRs, and for the pointwise prediction regions defined. Note although the last of these regions is not calibrated for simultaneous coverage, evaluating its simultaneous prediction performance serves as a reference point, given it is the most popular method of computing uncertainty quantification in FA (and mixed) models in the literature and software.

We simulated data from the FA model defined in \eqref{eq:vectorised-FR}, setting the number of latent variables to $m = 2$
and the number of covariates to $q=3$. For the latter, we constructed the covariate vectors as $\bm x_i=(1,x_{i2},x_{i3})^\top$
where $x_{ir}$ were simulated independently from a standard normal distribution for $r=2,3$. 
Next, the elements of the $pq$-vector of regression coefficients $\bm{\beta}$ were generated independently from a standard normal distribution, while for the
loading matrix $\bm{\Lambda}$, we first simulated all its entries independently from a standard normal distribution and then imposed corner constraints by setting $\lambda_{jk} = 0$ for all $j<k$ and $\lambda_{jk} \leftarrow |\lambda_{jk}|$ for $j=k$.
Finally, for the uniqueness matrix $\bm\Psi$, we defined the proportion of variance explained in the FA model as $\mathrm{VE} = \mathrm{tr}(\bm\Lambda\bm\Lambda^\top+\bm\Psi)^{-1}\mathrm{tr}(\bm\Lambda\bm\Lambda^\top)$, and then set $\psi_{jj}=(\mathrm{VE}^{-1} - 1)(\bm\Lambda\bm\Lambda^\top)_{jj},$ for $ j=1,\ldots,p$ after generating the loading matrix.
Higher values of VE correspond to stronger common-factor signals in the FA model.
We considered sample sizes $n\in\{50,100,200,500,1000\}$, the number of responses $p\in\{10,20\}$, and the variance explained $\mathrm{VE}\in\{0.3,0.5,0.7,0.9\}$, and for each combination of these simulated $I=1000$ datasets. Note $\bm\Lambda$, $\bm B$, and $\bm X$ are held fixed across each simulation setting, while new latent factors $\bm{f}$ and errors $\bm{\epsilon}$ are simulated for each dataset.

We set our the inferential target to be the collection of bivariate factor scores. That is, $d =m= 2$ and $\bm\mu_{0,l} = (f_{0,l1}, f_{0,l2})^\top$ for $l=1, 2, \dots, L$ and $L=n$. In terms of the multidimensional general mixed parameter formulation in \eqref{eq:def-multid-gmp}, this corresponds to setting $\bm C_l = \bm 0_{pq\times 2}$ and $\bm K_l$ to a sparse matrix with
two rows that are zero everywhere, except for a one at the $l$-th entry of the first row, and a one at the $(n+l)$-th entry of the second row.

For the bootstrap SPRs, we used $B=1000$ parametric bootstrap samples, and similarly for the Monte-Carlo SPRs we used $S=1000$ multivariate normal samples. 
Furthermore, for all four procedures compared we used $\bm G_{1,l}(\widehat{\bm\theta})$ as our estimate of the PMSE covariance matrix: recall that for both the bootstrap and Monte-Carlo procedures, we expect $\bm G_{1,l}(\widehat{\bm\theta})$ to offer more stable finite-sample performance of the SPRs, but to lead to undercoverage for Bonferroni SPRs. In SM~\ref{SM-sec:supplement-simres}, we present additional simulation results based on using $\widehat{\bm\Sigma}^{(\mathrm{full})}_l(\widehat{\bm \mu}_l) = \bm G_{1,l}(\widehat{\bm\theta})+\bm G_{2,l}(\widehat{\bm\theta})$ as our estimate of the PMSE covariance matrix, which confirm our expectations. 
All prediction regions were constructed at the nominal simultaneous coverage level $1-\alpha=0.95$.

\subsection{Simulation results} \label{subsec:simresults}

\begin{figure}[p]
\centering
\includegraphics[width=0.95\linewidth,
    trim=0 0.5cm 0 0, clip
    ]{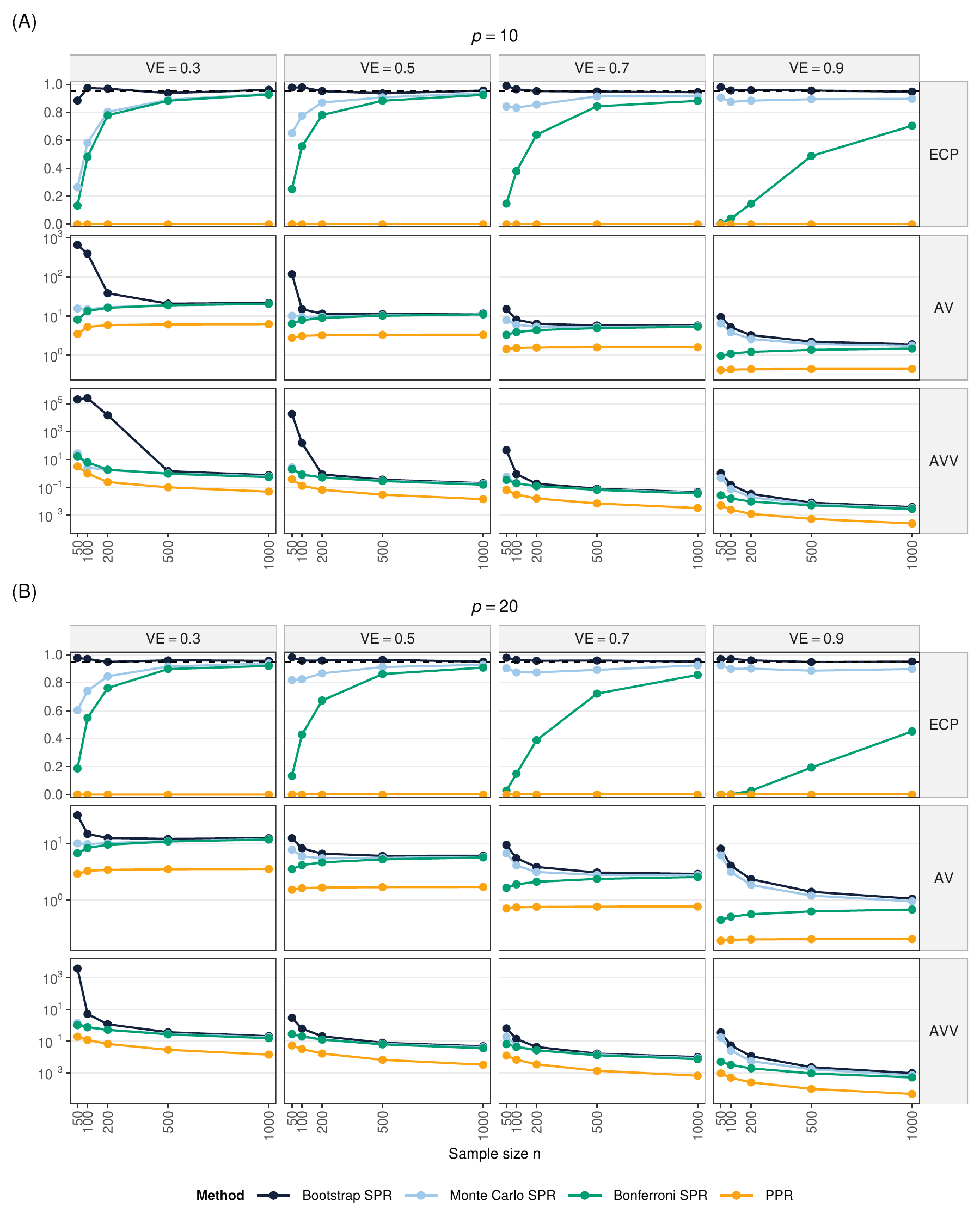}
\caption{
Comparison of four prediction regions methods when $p=10$ (Panel~A)  and $p=20$ (Panel~B), across different values of VE and sample size $n$. Metrics reported include the empirical simultaneous coverage probability (ECP), average volume (AV), and average variance of volume (AVV). The dashed horizontal line in the ECP panels is the nominal simultaneous coverage level $0.95$.}
\label{Fig:SPR-simulation-summary-G1}
\end{figure}

Figure~\ref{Fig:SPR-simulation-summary-G1} summarises the simulation results, with detailed numerical values given in SM~\ref{SM-sec:supplement-simres}.
Overall, the proposed bootstrap SPR produced simultaneous empirical coverage probabilities closest to the nominal level across almost all settings. While it was slightly conservative in some small-sample cases, particularly when the factor signal VE was weak, this conservativeness decreased as $n$ increased. The Monte-Carlo SPR also improved with sample size, although it tended to undercover more noticeably in smaller samples
In contrast, the Bonferroni SPR was less reliable: although in some low VE settings its performance improved as $n$ increased, it exhibited substantial undercoverage especially when the common-factor signal is strong. We do acknowledge however that this undercoverage was largely driven by the usage of only the leading term in constructing the estimated PMSE; in SM~\ref{SM-sec:supplement-simres}, we present results based on using $\widehat{\bm\Sigma}^{(\mathrm{full})}_l(\widehat{\bm \mu}_l)$ which showed much improved performance for Bonferroni SPRs, although they were still generally worse than the bootstrap SPRs. Unsurprisingly, the PPRs presented simultaneous coverage close to zero across almost all settings; this was consistent with the fact that its regions were substantially smaller in average volume compared with the three SPR methods.
For all methods, the decreasing values of $\mathrm{AVV}$ as $n$ increased showed that the estimated region volumes generally become more stable with greater sample size as expected. 

Comparing $p=10$ with $p=20$ (Panels A and B in Figure \ref{Fig:SPR-simulation-summary-G1}), we see that the overall trends are largely the same with bootstrap SPRs again performing best overall, although the regions for all methods were generally smaller and more stable at larger $p$. 
The was particularly visible in the average volume panels, where the low-VE settings lead to much larger regions when $p=10$ compared with $p=20$. 
Taken together, these results support the use of the bootstrap SPR as a  reliable finite-sample procedure for performing simultaneous prediction in FA models.

\section{Application to Coffee bean quality}
\label{sec:applications}

\begin{figure}[htb]
\centering
\includegraphics[width=\textwidth]{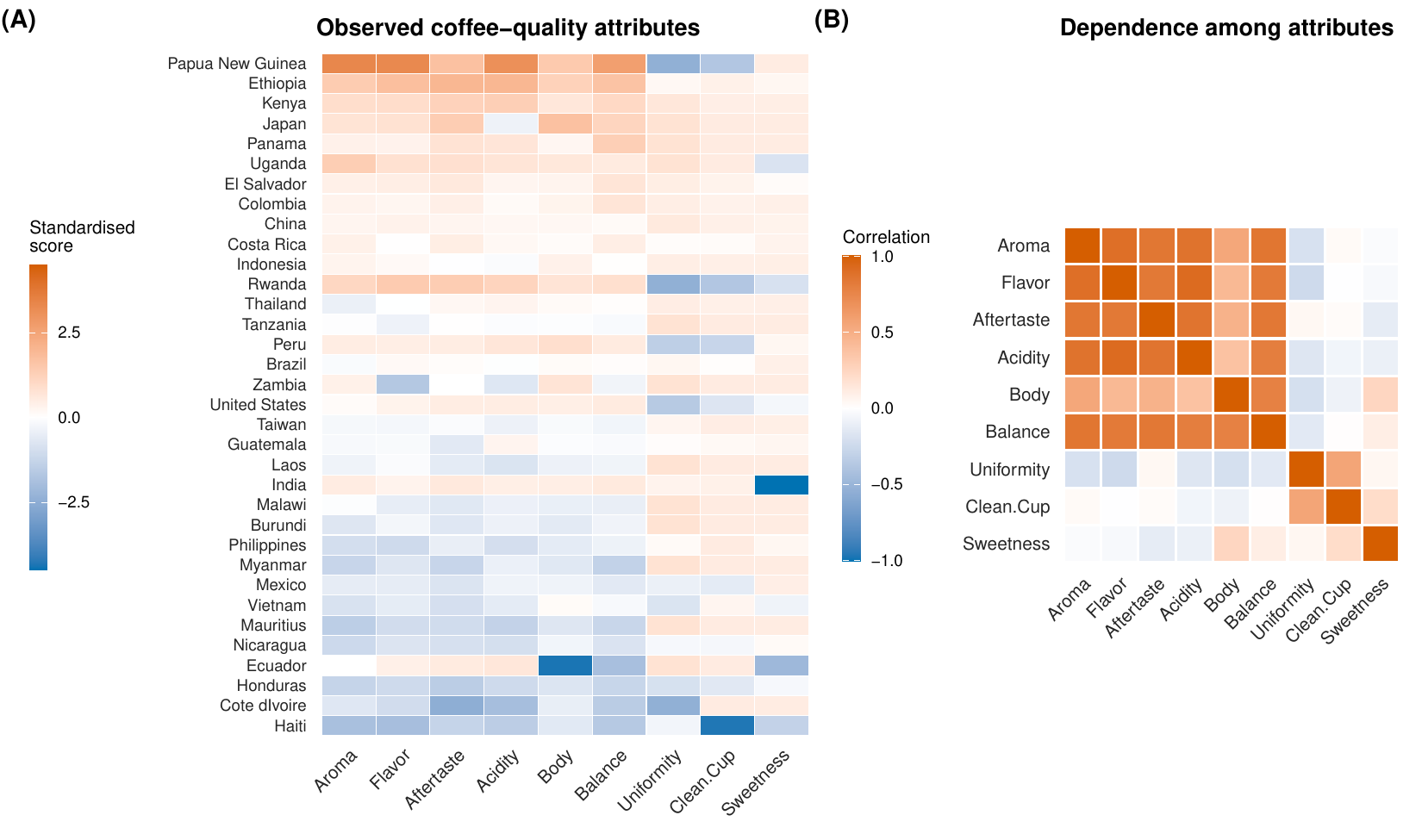}
\caption{
Visualisation of the coffee bean quality dataset. Panel~A presents the $p = 9$ (standardised) sensory attribute ratings for each of the $n = 34$ countries, ordered by their average ratings. Panel~B presents the sample correlation matrix of the nine attributes.
}
\label{Fig:visual_cof_ratings}
\end{figure}

We apply bootstrap SPRs to perform simultaneous prediction on a coffee bean quality dataset, originally sourced from the Coffee Quality Institute database \citep{coffeequalityinstitute} and publicly available from \url{https://www.kaggle.com/datasets/adampq/coffee-quality-with-locations-of-origin?resource=download}. The data comprise ratings of coffee beans from $n=34$ countries  across $p=9$ sensory attributes: Aroma, Flavor, Aftertaste, Acidity, Body, Balance, Uniformity, Clean Cup, and Sweetness. Note the ratings are on a continuous numerical scale between 0 and 10,  
but for illustrative purposes we will assume they are normally distributed and apply the FA model \eqref{eq:classic-FR}.
Figure~\ref{Fig:visual_cof_ratings} presents an exploratory overview of the data, showing that the nine attributes vary substantially  across countries. Furthermore, 
Aroma, Flavor, Aftertaste, Acidity, Body, and Balance are all strongly positively correlated, while Uniformity, Clean Cup, and Sweetness form a second positively correlated group. 

\begin{figure}[p]
\centering
\includegraphics[width=\textwidth]{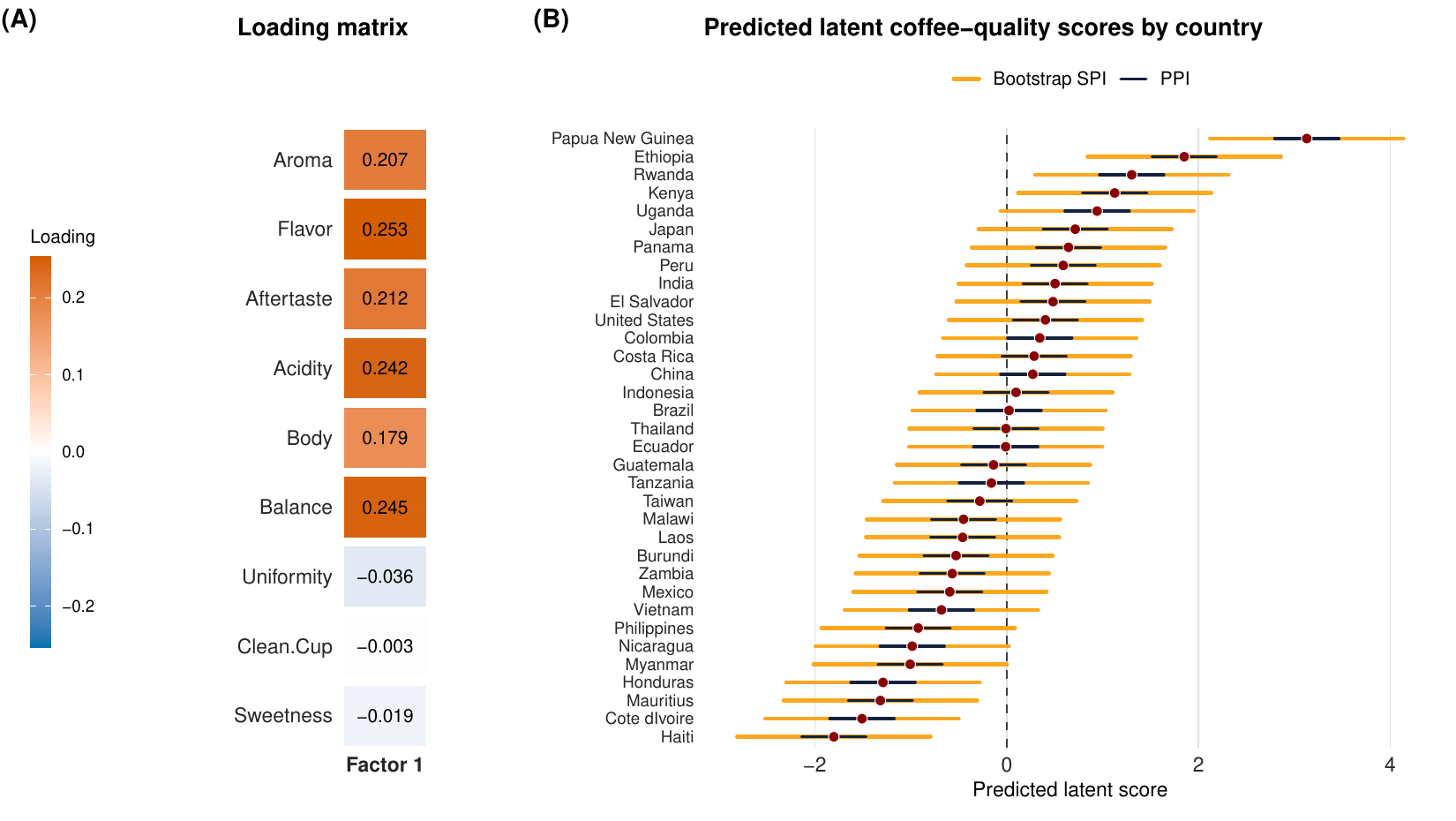}
\caption{
Estimated factor loadings and predicted latent coffee-quality scores under the $m = 1$ FA model. Panel~A presents the estimated loading matrix, while Panel~B depicts the country-specific predicted factor scores, ordered by their point estimates, together with bootstrap SPRs as well as PPRs (noting the regions reduce to intervals here).
}
\label{Fig:visual_cof_onefactor}
\end{figure}

We begin by fitting a $m = 1$ factor FA model, including only an intercept in $\bm{x}_i$ since no country-level covariates were available. The estimated VE level was $\widehat{\mathrm{VE}} \approx 0.3921$. Next, we set the general mixed parameter to be the latent factor $f_{i}$ for all individuals $i=1,\dots, n$, the simultaneous nominal level to be $0.95$, and constructed bootstrap SPRs using $B=10,000$ bootstrap samples.  
The resulting are shown in Figure~\ref{Fig:visual_cof_onefactor}, from which we see that, consistent with the exploratory analysis, the estimated $\bm{\Lambda}$ is dominated by positive loadings for Aroma, Flavor, Aftertaste, Acidity, Body, and Balance.
The predicted country-level factor scores $\widehat{f}_i$, which can thus be interpreted as a latent rating  primarily reflecting flavour- and balance-related preferences, are shown in the right panel,  
together with associated bootstrap SPRs as well as PPRs (noting these regions are intervals because we have a single factor, $m = 1$). 
Unsurprisingly, the bootstrap SPRs were substantially wider than their pointwise counterparts because they adjusted for simultaneous prediction across all countries. 
For instance, the PPRs for Papua New Guinea and Ethiopia did not overlap, with the former having the higher predicted factor score indicative of higher coffee quality. However, their corresponding bootstrap SPRs \emph{did} overlap, and as such, the countries are deemed to not be statistically different from each other after accounting for the simultaneity of the prediction; see Section~\ref{sec:discussion} for further discussion of this approach to pairwise comparisons based overlapping intervals.

\begin{figure}[p]
\centering
\includegraphics[width=\textwidth]{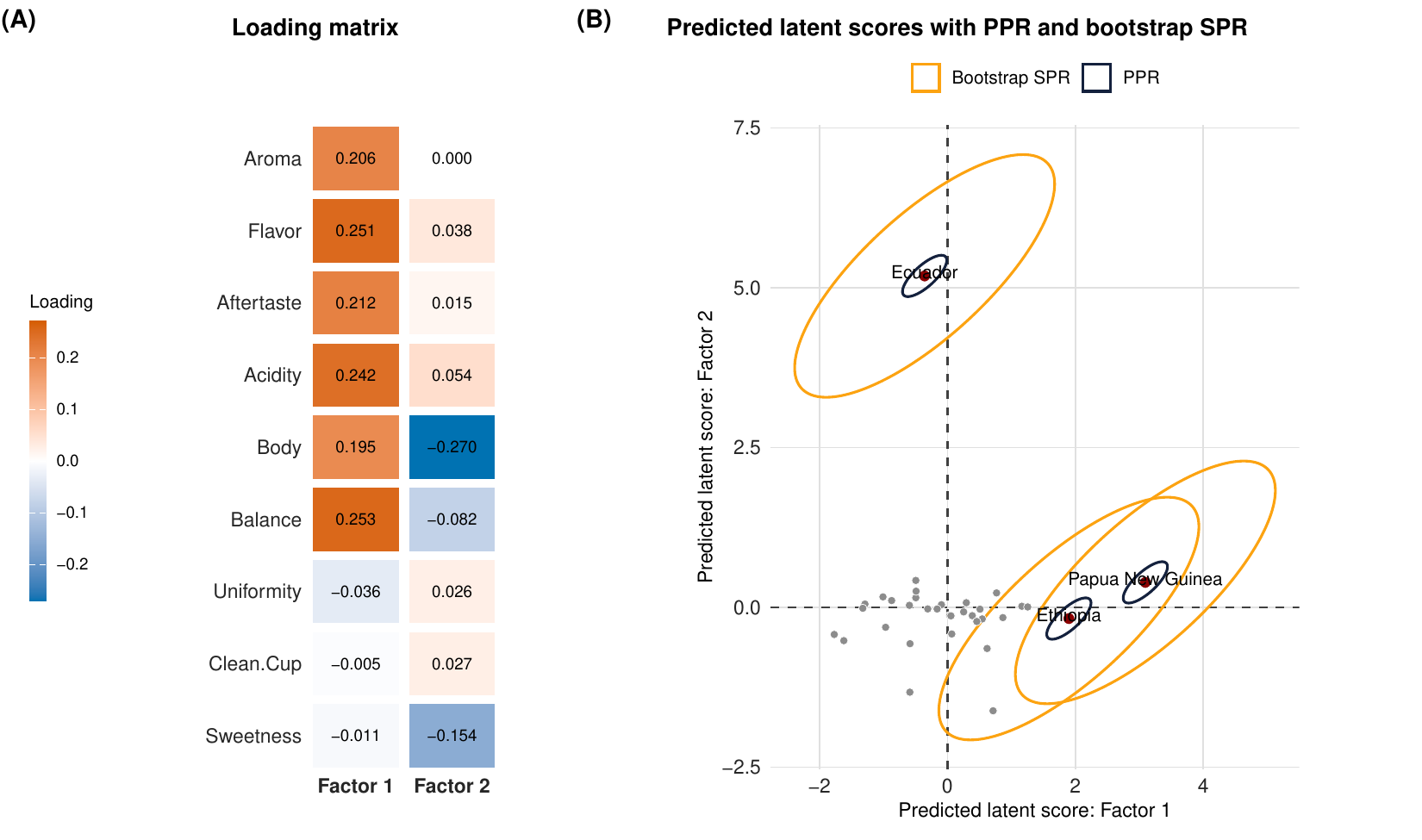}
\caption{
Estimated factor loadings and predicted latent coffee-quality scores under the $m = 2$ FA model. Panel~A presents the estimated loading matrix, while Panel~B depicts the country-specific predicted factor scores, together with bivariate bootstrap SPRs as well as PPRs for Papua New Guinea, Ethiopia, and Ecuador.
}
\label{Fig:visual_cof_joint}
\end{figure}

Next, we next fitted a $m = 2$ factor FA model and constructed prediction regions for the bivariate factor scores $(\widehat{f}_{i1}, \widehat{f}_{i2})^\top$ of each country.  The estimated VE level was $\widehat{\mathrm{VE}} \approx 0.5437$. The nominal level, number of bootstraps, and the choice of PMSE were set to be the same as in the $m=1$ case. The results are shown in Figure~\ref{Fig:visual_cof_joint}. Interpreting the loading matrix first, we see that the first factor is similar to the one-factor model in Figure \ref{Fig:visual_cof_onefactor} with positive loadings on Aroma, Flavour, Aftertaste, Acidity, Body, and Balance. The second factor is dominated by negative loadings on Body and Sweetness, suggesting a secondary axis in the sensory profile.
The right panel of Figure~\ref{Fig:visual_cof_joint} displays the predicted two-dimensional latent scores, along with PPRs and bootstrap SPRs shown for Papua New Guinea, Ethiopia, and Ecuador; these countries are chosen to avoid visual clutter and to exemplify some of the main trends. In particular, Papua New Guinea and Ethiopia were separated from the main cluster along the first factor, reflecting high latent sensory quality, whereas Ecuador was more clearly separated from the main cluster along the second factor.
More importantly, the bootstrap SPRs were again noticeably larger than the corresponding PPRs, e.g., 
the PPRs for Papua New Guinea and Ethiopia were well separated, but their bootstrap SPRs overlapped substantially. Thus, the simultaneous regions did not statistically distinguish between their bivariate latent coffee-quality profiles after accounting for simultaneous prediction. On the other hand, Ecuador remained clearly separated along the second factor even under the bootstrap SPR.

\section{Discussion}
\label{sec:discussion}

In this paper, we investigated simultaneous prediction procedures for multidimensional general mixed parameters in factor analytic models. In contrast to the standard conventional pointwise prediction regions, the proposed bootstrap SPRs control the probability that all targets are covered simultaneously, thus ensuring inference for factors scores across all observational units, say, is properly calibrated. 
Simulation results demonstrate the proposed bootstrap SPRs achieve reliable finite-sample simultaneous coverage, even when the sample size is small and/or the common-factor signal is weak, while alternative simpler methods based on Monte-Carlo simulation and a Bonferroni correction were less effective in general.
Simulation results also confirm that pointwise prediction regions produce simultaneous coverage far below the nominal level.
An application to a dataset on ratings of coffee bean quality illustrates how simultaneous prediction can potentially change the interpretation and conclusions drawn from FA models.
Note that in the application, we declared two countries as being statistically different from each other if their simultaneous prediction regions did not overlap. This is a somewhat conservative approach however, and more explicit hypothesis testing of pairwise differences between latent scores could instead be employed. Importantly, it is straightforward to extent the proposed bootstrap SPRs to perform such pairwise difference testing; see our discussion below \eqref{eq:def-multid-gmp} as well as \citet{reluga2023simultaneous}.

While the asymptotic developments in this paper focused on FA models with the number of general mixed parameters $L$ growing potentially faster than $n$, the dimensions $m$ and $q$ were kept fixed. Future research could extend the theory to settings in covariates and or factors grows with sample size \citep{li2017determining}. An immediate practical challenge for this high-dimensional case though is computation, as fitting let alone performing the parametric bootstrap on FA models may become computationally burdensome and require alternate approximation or optimisation strategies to manage this. The case of $m$ large also presents the issue how to select the number of latent variables, and future work could perhaps examine the theoretical and methodological impact of incorporating information criteria or regularisation, say, to select $m$ for simultaneous prediction \citep{hui2018order,ma2023generalized}. Besides, non-normal factor distributions \citep{zhang2014robust} offer a natural direction for extending the bootstrap construction and its validity theory. Finally, analogous to \citep{reluga2023simultaneousGLMM}, extensions to the case of non-normal responses and generalised linear latent variable models or some variation thereof \citep{skrondal2004generalized} presents an important avenue of future investigation, although issues around computational scalability are even more prominent here \citep[see][and references therein for examples of recent work on this]{kidzinski2022generalized,korhonen2023fast}.

\section*{Supplementary Materials}
The Supplementary Material (SM) 
contains auxiliary lemmas and complete proofs of the theoretical results; formal algorithms for the bootstrap and Monte-Carlo SPRs, including the bootstrap factor-alignment procedure; additional discussion of the choice of PMSE approximation and the relationship between Bonferroni and Gumbel calibration; and supplementary simulation results.
\par
\section*{Acknowledgements}
This projects was supported by an Australian Research Council Discovery Project DP230101908.

\bibliographystyle{chicago}
\bibliography{references}

\clearpage

\begin{center}
{\large\bfseries Supplementary Material for Wang et al., Simultaneous Prediction}\\[2pt]
{\large\bfseries for Latent Variables in Factor Analytic Models}
\end{center}

\vspace{.5cm} 
\noindent
This Supplementary Material contains the technical details and additional results supporting the main paper. 
\begin{itemize}
    \item Section~\ref{SM-sec:theory-and-proofs} presents the regularity conditions and proofs of the theoretical results. Specifically, 
    \begin{itemize}
        \item Section~\ref{SM-sec:lemmas} states the supporting lemmas used in the subsequent theoretical developments; 
        \item Section~\ref{SM-sec:prop1} presents Proposition~\ref{SM-prop:normal-approx} together with its proof; 
        \item Section~\ref{SM-sec:proof-thm1} provides the proof of Theorem~\ref{thm:bootstrap-consistency} from the main text;
        \item Section~\ref{SM-sec:proof-prop2} provides the proof of Proposition~\ref{prop:distribution-of-delta} from the main text; and 
        \item Section~\ref{SM-sec:supplement-proof-of-lemma} contains the deferred proofs of the lemmas stated in Section~\ref{SM-sec:lemmas}.
    \end{itemize}

    \item Section~\ref{SM-sec:supp-algorithms} presents additional methodological details, including the bootstrap alignment procedure, the choice of prediction mean squared error approximation, the connection between Bonferroni and Gumbel calibration, and detailed algorithms for constructing simultaneous prediction regions. 
    \item Section~\ref{SM-sec:supplement-simres} provides additional detailed simulation results.
\end{itemize}

\par

\setcounter{section}{0}
\setcounter{equation}{0}
\setcounter{table}{0}
\setcounter{figure}{0}
\setcounter{lemma}{0}
\setcounter{proposition}{0}
\setcounter{remark}{0}
\setcounter{algorithm}{0}
\def\theequation{S\arabic{section}.\arabic{equation}}
\def\thesection{S\arabic{section}}
\renewcommand{\theproposition}{S\arabic{proposition}}
\renewcommand{\thelemma}{S\arabic{lemma}}

\section{Theory and proofs}
\label{SM-sec:theory-and-proofs}

Let $\zeta_{\min}(\bm{M})$ and $\zeta_{\max}(\bm{M})$ denote the minimum and maximum eigenvalues of a symmetric matrix $\bm{M}$, respectively, and assume that the number of factors $m$ and the dimension of the general mixed parameter $d$ are fixed. Furthermore, assume that the elements in $\bm{\theta}$ corresponding the diagonal elements of $\bm{\Lambda}$ have been reparameterised, e.g., via a log transformation, such that they are now unconstrained. For ease of notation though, we will not make this reparametrisation explicit in the theoretical developments below. Additionally, throughout this section, $\bm R^{(k)}_{n^{-a}}$ denotes a generic conformable random matrix whose elements are of order $O_p(n^{-a})$; $k$ distinguishes remainders.

For convenience, we restate the regularity conditions provided in Section 3.1 in the main text.

\begin{enumerate}[label=(A\arabic*), ref=A\arabic*]
    \item The true parameter vector $(\bm{\beta}_0^\top, \bm{\theta}_0^\top)^\top$ is an interior point of the parameter space $\bm{\Theta} \subset \mathbb{R}^{pq + h}$, where $h= pm-2^{-1} m(m-1) + p$.
    \label{SM-regcond:true-para}
    \item There exist constants $0< n_0 < \infty$, $0< C_0<\infty$ and $1 <C_1 < \infty$ such that for every $n \ge n_0$ and $i = 1,\ldots,n$, it holds that $\|\bm{x}_i\|_\infty \le C_0$, and $C_1^{-1} \le \zeta_{\min}(n^{-1} \sum_{i=1}^n \bm{x}_i\bm{x}_i^\top) \le \zeta_{\max}(n^{-1}\sum_{i=1}^n \bm{x}_i\bm{x}_i^\top) \le C_1$. \label{SM-regcond:X}
    \item Write $\bm{K}_l^ \top = \left[\bm{K}_{l1} \; \bm{K}_{l2} \; \cdots \; \bm{K}_{ln} \right]$ where $\bm{K}_{l1}$ is a $d \times m$ matrix, and define
    $\mathcal{S}_l = \left\{i: \bm{K}_{li} \neq \bm{0}_{d\times m} \right\}$. Then there exist sufficiently large positive constants $S_K, C_K$ and $C_L$ such that $\sup_{1 \leq l \leq L} |\mathcal{S}_l| \leq S_K$ and $\sup_{1 \leq l \leq L} \sup_{1 \leq i \leq n} \|\bm{K}_{li}\| \leq C_K$. Furthermore, $\max_{1\leq l \leq L}\|\bm{C}_l\| = C_L$. \label{SM-regcond:designed-mat}
    \item There exists an open subset $\omega$ of $\mathbb{R}^{h}$ containing $\bm\theta_0$ such that, for all $j = 1,\ldots,h$, it holds that $\sup_{\bm\theta \in \omega} \max_{1\leq l\leq L} \| \partial / \partial\theta_j (\bm I_p\otimes\bm X)^\top\bm O_l(\bm\theta) \| \le C_2,
    $
    for some sufficiently large positive constant $C_2$.
    \label{SM-regcond:D-mat}
    \item  In the same locally identified parameterisation, $\|\widehat{\bm{\theta}} - \bm{\theta}_0\| = O_p(n^{-1/2})$ and $\| \widehat{\bm\theta}^* - \widehat{\bm\theta}\| = O_{p^*}(n^{-1/2})$ in probability under the original sampling law. \label{SM-regcond:MLE}
    \item There exist constants $1 < \kappa_1, \kappa_2 < \infty$ such that $\kappa_1^{-1} \le \zeta_{\min}\{\bm{G}_{1,l}(\bm{\theta}_0)\} \le \zeta_{\max}\{\bm{G}_{1,l}(\bm{\theta}_0)\} \le \kappa_1$, for all $l = 1,2, \dots,L$, and $\kappa_2^{-1} \le \zeta_{\min}(\bm{V}_0) \le \zeta_{\max}(\bm{V}_0) \le \kappa_2$. \label{SM-regcond:PMSE}
\end{enumerate}

\subsection{Lemmas}
\label{SM-sec:lemmas}

To facilitate proofs of propositions and theorems, we propose all lemmas used here and defer their proofs to Section~\ref{SM-sec:supplement-proof-of-lemma}. 

\begin{lemma}
\label{SM-lemma:variance-component-approx}
    Assume Conditions \ref{SM-regcond:true-para}--\ref{SM-regcond:PMSE} hold. Then the estimated variance matrix of $\bm{y}_i$ for any $i \in [n]$ admits the following expansion $$
    \widehat{\bm{V}}^{-1} = \bm{V}_0^{-1} -\bm{V}_0^{-1} \bm{A}(\widehat{\bm{\theta}}, \bm{\theta}_0) \bm{V}_0^{-1} + \ordR{-1},
    $$
    where $\bm{A}(\widehat{\bm{\theta}}, \bm{\theta}_0) = (\widehat{\bm{\Lambda}} - \bm{\Lambda}_0) \bm{\Lambda}_0^\top + \bm{\Lambda}_0 (\widehat{\bm{\Lambda}} - \bm{\Lambda}_0)^\top + (\widehat{\bm{\Psi}} - \bm{\Psi}_0)$.
\end{lemma}

\begin{lemma}
\label{SM-lemma:order_of_G1andG2}
    Assume Conditions \ref{SM-regcond:true-para}--\ref{SM-regcond:PMSE} hold. Then 
    $$
    \max_{1 \leq l \leq L} \|\bm G_{1,l}(\bm{\widehat{\theta}}) - \bm G_{1,l}(\bm\theta_0)\| = O_p(n^{-1/2})
    $$
    and
    $$
    \max_{1 \leq l \leq L} \|\bm G_{2,l}(\bm{\widehat{\theta}})\| = O_p(n^{-1}).
    $$
\end{lemma}

\begin{lemma}
\label{SM-lemma:order-leader}
    Assume Conditions~\ref{SM-regcond:true-para}--\ref{SM-regcond:PMSE} hold. Let
    $\bm Z_l(\bm\theta_0)=\bm G_{1,l}^{-1/2}(\bm\theta_0)\bm K_l^\top\{\check{\bm f}(\bm\theta_0)-\bm f\}$
    be the oracle Gaussian leading term, where
    $\check{\bm f}(\bm\theta_0):=\{(\bm\Lambda_0^\top\bm V_0^{-1})\otimes\bm I_n\}\{\bm y-(\bm I_p\otimes\bm X)\bm\beta_0\}$.
    Then
    $$
    \max_{1\le l\le L}\|\bm Z_l(\bm\theta_0)\|=O_p(\sqrt{\log L}),
    \qquad
    \mathbb E\left\{\max_{1\le l\le L}\|\bm Z_l(\bm\theta_0)\|\right\}=O(\sqrt{\log L}).
    $$
\end{lemma}

\begin{lemma}
    Assume Conditions \ref{SM-regcond:true-para}--\ref{SM-regcond:PMSE} hold and the size of the collection of general mixed parameter targets satisfies $\log L = o(\sqrt{n})$. Define $\bm{T}_{l,n} = \widehat{\bm{\Sigma}}_l^{-1/2} \{\widehat{\bm{\mu}}_l(\widehat{\bm{\theta}}) - \bm{\mu}_l\}$. Then 
    \begin{equation*}
        \max_{1\leq l \leq L} \|\bm{T}_{l,n} - \bm{Z}_l \| = O_p \left(\sqrt{\frac{\log L}{n}}. \right)
    \end{equation*}
    \label{SM-lemma:order-reminder}
\end{lemma}

\begin{lemma}[Theorem 2.1 in \cite{chernozhukov2014anti}]
\label{SM-lemma:thm2.1-CCK}
    Let $\{X_t:t\in\mathcal T\}$ be a separable Gaussian process satisfying $\mathbb E(X_t)=0$ and $\mathbb E(X_t^2)=1$ for all $t\in\mathcal T$, and suppose $\sup_{t\in\mathcal T}X_t<\infty$ almost surely. Define $a(X):=\mathbb E(\sup_{t\in\mathcal T}X_t)$. Then, for every $\epsilon\geq0$,
    $$
    \sup_{x\in\mathbb R}\mathbb P\left(\left|\sup_{t\in\mathcal T}X_t-x\right|\leq\epsilon\right)\leq4\epsilon\{a(X)+1\}.
    $$
\end{lemma}
\begin{remark}
A stochastic process $\{X_t:t\in\mathcal T\}$ is called separable if there exists a countable subset $\mathcal T_0\subseteq\mathcal T$ such that $\sup_{t\in\mathcal T}X_t = \sup_{t\in\mathcal T_0}X_t$
almost surely. In other words, although the index set $\mathcal T$ may be uncountable, the supremum of the process can be determined from a countable subset.
\end{remark}

\subsection{Proposition S1 and its proof}
\label{SM-sec:prop1}

\begin{proposition} \label{SM-prop:normal-approx}
    Assume Conditions \ref{SM-regcond:true-para}--\ref{SM-regcond:PMSE} hold. For each fixed target $l$, as $n \rightarrow \infty$, we have
    \begin{align*}
        \sup_{\bm{t}} |\mathbb{P}(\bm{T}_{l,n} \le \bm{t}) - \Phi_d(\bm{t})| \rightarrow 0, \quad \forall l \in [L]    
    \end{align*}
    where inequalities are componentwise, $\Phi_d(\cdot)$ denotes the CDF of a $d$-variate standard normal distribution.
\end{proposition}

\begin{proof}
    Write 
    \begin{align*}
        \widehat{\bm{\Sigma}}_l^{-1/2}(\widehat{\bm{\theta}})\left\{\widehat{\bm{\mu}}_l(\widehat{\bm{\theta}}) - \bm{\mu}_l \right\} = \left\{\bm{G}_{1,l}(\widehat{\bm{\theta}}) + \bm{G}_{2,l} (\widehat{\bm{\theta}})\right\}^{-1/2} \left\{ \bm{C}_l^\top(\widehat{\bm{\beta}} - \bm{\beta}) +\bm{K}_l^\top (\widehat{\bm{f}}(\widehat{\bm{\theta}}) - \bm{f}) \right\}.
    \end{align*}
    First, we deal with $\bm{K}_l^\top \{\widehat{\bm{f}}(\widehat{\bm{\theta}}) - \bm{f}\}$. Consider $\widehat{\bm{f}}(\widehat{\bm{\theta}}) - \widehat{\bm{f}}(\bm{\bm{\theta}_0})$, where $\bm{\theta}_0$ denotes the true variance components. Then we have
    \begin{align*}
        \widehat{\bm{f}}(\widehat{\bm{\theta}}) - \widehat{\bm{f}}(\bm{\bm{\theta}_0}) 
        &= \left(\widehat{\bm{\Lambda}}^\top \widehat{\bm{V}}^{-1} \otimes \bm{I}_n\right) \left\{\bm{y} - (\bm{I}_p \otimes \bm{X})\widehat{\bm{\beta}}\right\} \\
        & \qquad - \left(\bm{\Lambda}_0^\top \bm{V}_0^{-1} \otimes \bm{I}_n\right) \left\{\bm{y} - (\bm{I}_p \otimes \bm{X})\widehat{\bm{\beta}}\right\}\\ 
        &= \left\{ \left(\widehat{\bm{\Lambda}}^\top \widehat{\bm{V}}^{-1} - \bm{\Lambda}_0^\top \bm{V}_0^{-1} \right) \otimes \bm{I}_n\right\} \left\{\bm{y} - (\bm{I}_p \otimes \bm{X})\widehat{\bm{\beta}}\right\} \\
        &= \left\{ \left(\widehat{\bm{\Lambda}}^\top \widehat{\bm{V}}^{-1} - \bm{\Lambda}_0^\top \bm{V}_0^{-1} \right) \otimes \bm{I}_n\right\} \left\{\bm{y} - (\bm{I}_p \otimes \bm{X})\bm{\beta}_0\right\} \\
        &\qquad + \left\{ \left(\widehat{\bm{\Lambda}}^\top \widehat{\bm{V}}^{-1} - \bm{\Lambda}_0^\top \bm{V}_0^{-1} \right) \otimes \bm{I}_n\right\} \left(\bm{I}_p \otimes \bm{X} \right) \left\{ \bm{\beta}_0 - \widehat{\bm{\beta}} \right\}, 
    \end{align*}
    where the second line uses the fact that the OLS estimator does not depend on $\bm{\theta}$. By Assumption \ref{SM-regcond:MLE}, we know $\|\widehat{\bm{\theta}} - \bm{\theta}_0\| = O_p(n^{-1/2})$ and $\|\widehat{\bm{\beta}} - \bm{\beta}_0\| = O_p(n^{-1/2})$ and therefore
    \begin{align*}
        \widehat{\bm{f}}(\widehat{\bm{\theta}}) - \widehat{\bm{f}}(\bm{\bm{\theta}_0}) 
        &= \left\{\left(\widehat{\bm{\Lambda}}^\top \widehat{\bm{V}}^{-1} - \bm{\Lambda}_0^\top \bm{V}_0^{-1} \right) \otimes \bm{I}_n\right\} \left\{\bm{y} - (\bm{I}_p \otimes \bm{X})\bm{\beta}_0\right\} + \ordR{-1} \\
        &= \left\{(\widehat{\bm{\Lambda}} - \bm{\Lambda}_0)^\top \bm{V}_0^{-1} \otimes \bm{I}_n \right\} \left\{\bm{y} - (\bm{I}_p \otimes \bm{X})\bm{\beta}_0\right\} \\
        & \qquad - \left\{\bm{\Lambda}_0^\top \bm{V}_0^{-1} \bm{A}(\widehat{\bm{\theta}}, \bm{\theta}_0) \bm{V}_0^{-1} \otimes \bm{I}_n \right\}\left\{\bm{y} - (\bm{I}_p \otimes \bm{X})\bm{\beta}_0\right\} + \ordR{-1},
    \end{align*}
    where the second line follows from Lemma \ref{SM-lemma:variance-component-approx}. Observing that $(\bm{V}_0^{-1} \otimes \bm{I}_n) \{\bm{y} - (\bm{I}_p \otimes \bm{X})\bm{\beta}_0\} \sim \mathcal{N}_{np} \left(\bm{0}_{np}, \bm{V}_0^{-1} \otimes \bm{I}_n \right)$, then each selected unit contributes a Gaussian vector of bounded variance. With $p,m$ fixed, Assumption~\ref{SM-regcond:designed-mat} selects a bounded number of such units with bounded coefficients. Combining this with Assumption~\ref{SM-regcond:MLE} and the preceding expansion gives
    \begin{equation}
        \| \bm{K}_l^\top \widehat{\bm{f}}(\widehat{\bm{\theta}}) - \bm{K}_l^\top \widehat{\bm{f}}(\bm{\theta}_0) \|_F = O_p(n^{-1/2}).
    \end{equation}

    We now turn to the matrix $\bm{G}_{1,l}(\widehat{\bm{\theta}})$, by the Woodbury identity, we have
    $$\bm{G}_{1,l}(\widehat{\bm{\theta}}) := \bm{K}_l^\top \left\{ \bm{I}_{mn} - (\widehat{\bm{\Lambda}}^\top \widehat{\bm{V}}^{-1}\widehat{\bm{\Lambda}}) \otimes \bm{I}_n \right\}\bm{K}_l = \bm{K}_l^\top \left\{ \bm{I}_{mn} + (\widehat{\bm{\Lambda}}^\top \widehat{\bm{\Psi}}^{-1}\widehat{\bm{\Lambda}}) \otimes \bm{I}_n \right\}^{-1}\bm{K}_l.
    $$ 
    From a Taylor expansion, we have $\widehat{\psi}^{-1}_{jj} = \psi^{-1}_{0,jj} - \psi^{-2}_{0,jj} (\widehat{\psi}_{jj} - \psi_{0,jj}) + O_p(n^{-1})$. Thus applying this for all $j = 1,\ldots,p$, we obtain 
    \begin{align*}
    \bm{\widehat{\Psi}}^{-1} = \bm{\Psi}^{-1}_0 - \bm{\Psi}^{-1}_0 (\bm{\widehat{\Psi}} - \bm{\Psi}_0) \bm{\Psi}^{-1}_0 + \ordR{-1}.
    \end{align*}
    It follows that
    \begin{align*}
    \widehat{\bm{\Lambda}}^\top \widehat{\bm{\Psi}}^{-1} \widehat{\bm{\Lambda}} &= (\bm{\Lambda}_0 + \widehat{\bm{\Lambda}} - \bm{\Lambda}_0)^\top \left\{\bm{\Psi}^{-1}_0 - \bm{\Psi}^{-1}_0 (\bm{\widehat{\Psi}} - \bm{\Psi}_0) \bm{\Psi}^{-1}_0 + \ordR{-1}\right\} (\bm{\Lambda}_0 + \widehat{\bm{\Lambda}} - \bm{\Lambda}_0) \\
    &= \bm{\Lambda}_0^\top \bm{\Psi}^{-1}_0 \bm{\Lambda}_0 + \bm{\Lambda}_0^\top \bm{\Psi}^{-1}_0 (\widehat{\bm{\Lambda}} - \bm{\Lambda}_0) + (\widehat{\bm{\Lambda}} - \bm{\Lambda}_0)^\top  \bm{\Psi}^{-1}_0 \bm{\Lambda}_0 \\
    &\qquad - \bm{\Lambda}_0^\top \bm{\Psi}^{-1}_0 (\bm{\widehat{\Psi}} - \bm{\Psi}_0) \bm{\Psi}^{-1}_0 \bm{\Lambda}_0 + \ordR{-1},
    \end{align*}
    where the last three displayed terms before the remainder are the linear perturbations in the loadings and uniquenesses. For $\bm{G}_{1,l}(\widehat{\bm{\theta}})$, we obtain
    \begin{align*}
    \bm{K}_l^\top &\left\{\bm{I}_{nm} + \left(\widehat{\bm{\Lambda}}^\top \widehat{\bm{\Psi}}^{-1} \widehat{\bm{\Lambda}}\right) \otimes \bm{I}_n \right\}^{-1}\bm{K}_l \\ 
    &= \bm{K}_l^\top \left\{\bm{I}_{mn} + \left(\bm{\Lambda}_0^\top \bm{\Psi}^{-1}_0 \bm{\Lambda}_0 + \widehat{\bm{\Lambda}}^\top \widehat{\bm{\Psi}}^{-1} \widehat{\bm{\Lambda}} - \bm{\Lambda}_0^\top \bm{\Psi}^{-1}_0 \bm{\Lambda}_0 \right) \otimes \bm{I}_n \right\}^{-1}\bm{K}_l \\
    &= \bm{K}_l^\top \left\{ \left(\bm{I}_{mn} + (\bm{\Lambda}_0^\top \bm{\Psi}^{-1}_0 \bm{\Lambda}_0) \otimes \bm{I}_n\right) + (\widehat{\bm{\Lambda}}^\top \widehat{\bm{\Psi}}^{-1} \widehat{\bm{\Lambda}} - \bm{\Lambda}_0^\top \bm{\Psi}^{-1}_0 \bm{\Lambda}_0)\otimes \bm I_n \right\}^{-1}\bm{K}_l \\
    &= \bm{G}_{1,l}(\bm{\theta}_0) - \bm{K}_l^\top\left\{\bm{I}_{mn} + (\bm{\Lambda}_0^\top \bm{\Psi}^{-1}_0 \bm{\Lambda}_0) \otimes \bm{I}_n\right\}^{-1}\\
    & \qquad \times \left\{(\widehat{\bm{\Lambda}}^\top \widehat{\bm{\Psi}}^{-1} \widehat{\bm{\Lambda}} - \bm{\Lambda}_0^\top \bm{\Psi}^{-1}_0 \bm{\Lambda}_0) \otimes\bm I_n \right\}\left(\bm{I}_{mn} + (\bm{\Lambda}_0^\top \bm{\Psi}^{-1}_0 \bm{\Lambda}_0) \otimes \bm{I}_n\right)^{-1}\bm{K}_l + \ordR{-1},
    \end{align*}
    where the last line follows in a similar vein to the expansion used in the proof of Lemma~\ref{SM-lemma:variance-component-approx}. Finally, observing that $\|\bm{\Lambda}_0^\top \bm{\Psi}^{-1}_0 (\widehat{\bm{\Lambda}} - \bm{\Lambda}_0)\| = O_p(n^{-1/2})$, $\| (\widehat{\bm{\Lambda}} - \bm{\Lambda}_0)^\top  \bm{\Psi}^{-1}_0 \bm{\Lambda}_0\| = O_p(n^{-1/2})$, and $\|\bm{\Lambda}_0^\top \bm{\Psi}^{-1}_0 (\bm{\widehat{\Psi}} - \bm{\Psi}_0) \bm{\Psi}^{-1}_0 \bm{\Lambda}_0\| = O_p(n^{-1/2})$, all of which can be shown via straightforward applications of the Cauchy-Schwartz inequality, then it follows that $\|\widehat{\bm{\Lambda}}^\top \widehat{\bm{\Psi}}^{-1} \widehat{\bm{\Lambda}} - \bm{\Lambda}_0^\top \bm{\Psi}^{-1}_0 \bm{\Lambda}_0\|_F = O_p(n^{-1/2})$ and hence
    \begin{equation} \label{SM-eq:G1-norm}
    \|\bm{G}_{1,l}(\widehat{\bm{\theta}}) - \bm{G}_{1,l}(\bm{\theta}_0)\|_F = O_p(n^{-1/2}). 
    \end{equation}

    \noindent Now consider 
    \begin{align*}
        \bm{K}_l^\top (\widehat{\bm{f}}(\widehat{\bm{\theta}}) - \bm{f}) = \bm{K}_l^\top\widehat{\bm{f}}(\widehat{\bm{\theta}}) - \bm{K}_l^\top\widehat{\bm{f}}(\bm{\theta}_0) + \bm{K}_l^\top\widehat{\bm{f}}(\bm{\theta}_0) - \bm{K}_l^\top\bm{f}.
    \end{align*}
    Under $\bm{\theta}_0$, we have $\bm{K}_l^\top\widehat{\bm{f}}(\bm{\theta}_0) - \bm{K}_l^\top\bm{f} = \bm{G}_{1,l}^{1/2}(\bm{\theta}_0) \mathcal{Z}_d +\ordR{-1/2}$ where $\mathcal{Z}_d \sim \mathcal{N}_d(\bm{0}_d, \bm{I}_d)$.
    Here $\mathcal Z_d=\bm Z_l(\bm\theta_0)$ from Lemma~\ref{SM-lemma:order-leader}; the additional remainder follows from
    \begin{align*}
    \bm K_l^\top\{\widehat{\bm f}(\bm\theta_0)-\check{\bm f}(\bm\theta_0)\}
    &=-\bm K_l^\top(\bm\Lambda_0^\top\bm V_0^{-1}\otimes\bm I_n)
      (\bm I_p\otimes\bm X)(\widehat{\bm\beta}-\bm\beta_0)=O_p(n^{-1/2}),
    \end{align*}
    by Assumptions~\ref{SM-regcond:X} and \ref{SM-regcond:designed-mat}.
    It follows that $\bm{K}_l^\top \{\widehat{\bm{f}}(\widehat{\bm{\theta}}) - \bm{f}\} = \bm{G}_{1,l}^{1/2}(\bm{\theta}_0) \mathcal{Z}_d + \ordR{-1/2}$, and furthermore we obtain
    \begin{align*}
        \bm{G}_{1,l}^{-1/2}(\widehat{\bm{\theta}})\left[\bm{K}_l^\top \{\widehat{\bm{f}}(\widehat{\bm{\theta}}) - \bm{f}\}\right] 
        &= \bm{G}_{1,l}^{-1/2}(\widehat{\bm{\theta}}) \bm{G}_{1,l}^{1/2}(\bm{\theta}_0) \mathcal{Z}_d + \ordR{-1/2} \\
        &= \mathcal{Z}_d + \{\bm{G}_{1,l}^{-1/2}(\widehat{\bm{\theta}}) - \bm{G}_{1,l}^{-1/2}(\bm{\theta}_0)\}\bm{G}_{1,l}^{1/2}(\bm{\theta}_0)\mathcal{Z}_d + \ordR{-1/2} \\
        &= \mathcal{Z}_d + \ordR{-1/2},
    \end{align*}
    where the third line follows from equation \eqref{SM-eq:G1-norm} and local Lipschitz continuity of the inverse square-root map under Assumption~\ref{SM-regcond:PMSE}.

Turning to $\bm{C}_l^\top(\widehat{\bm{\beta}} - \bm{\beta})$, then from Assumption \ref{SM-regcond:designed-mat} and the OLS rate above, we directly have $\bm{C}_l^\top(\widehat{\bm{\beta}} - \bm{\beta}) = \ordR{-1/2}$.

    Lastly, consider $\bm{G}_{2,l}(\widehat{\bm{\theta}}) := \{\bm{C}_l - (\bm{I}_p \otimes \bm{X}^\top)\bm{O}_l(\widehat{\bm{\theta}})\}^\top \bm{Q}(\widehat{\bm{\theta}}) \{\bm{C}_l - (\bm{I}_p \otimes \bm{X}^\top)\bm{O}_l(\widehat{\bm{\theta}})\}$, where $\bm{O}_l(\bm{\theta}) := (\bm{V}^{-1}\bm{\Lambda} \otimes \bm{I}_n)\bm{K}_l$ and $\bm{Q}(\bm{\theta}):= \bm{V} \otimes (\bm{X}^\top\bm{X})^{-1}$. The matrix $\bm Q(\widehat{\bm\theta})$ is of order $O_p(n^{-1})$ element-wise, since\
    \begin{align*}
        \bm{Q}(\widehat{\bm{\theta}}) = \widehat{\bm{V}} \otimes (\bm{X}^\top \bm{X})^{-1} = (\bm{V}_0 + \ordR{-1/2}) \otimes (\bm{X}^\top \bm{X})^{-1},
    \end{align*}
    according to Lemma \ref{SM-lemma:variance-component-approx} and Assumption \ref{SM-regcond:X}. For the factor $\bm C_l-(\bm I_p\otimes\bm X^\top)\bm O_l(\widehat{\bm\theta})$, observing that
    \begin{align*}
        \bm{O}_l(\widehat{\bm{\theta}}) - \bm{O}_l(\bm{\theta}_0)
        &= \left\{\bm{V}_0^{-1}(\widehat{\bm{\Lambda}} - \bm{\Lambda}_0) \otimes \bm{I}_n \right\} \bm{K}_l +  \left\{(\widehat{\bm{V}}^{-1} - \bm{V}_0^{-1})\widehat{\bm{\Lambda}} \otimes \bm{I}_n \right\} \bm{K}_l \\
        &= \left\{\bm{V}_0^{-1}(\widehat{\bm{\Lambda}} - \bm{\Lambda}_0) \otimes \bm{I}_n \right\} \bm{K}_l - \left\{\bm{V}_0^{-1}\bm{A}(\widehat{\bm{\theta}}, \bm{\theta}_0)\bm{V}_0^{-1} \bm{\Lambda}_0 \otimes \bm{I}_n \right\}\bm{K}_l + \ordR{-1},
    \end{align*}
    by Lemma \ref{SM-lemma:variance-component-approx}, then we have
    \begin{equation*}
        \|\bm{O}_l(\widehat{\bm{\theta}}) - \bm{O}_l(\bm{\theta}_0)\|_F = O_p(n^{-1/2}).
    \end{equation*}
    Thus, by Assumptions~\ref{SM-regcond:X}--\ref{SM-regcond:D-mat} (see the bounds in the proof of Lemma~\ref{SM-lemma:order_of_G1andG2}), this factor can be written as
    \begin{align*}
        \bm{C}_l - (\bm{I}_p \otimes \bm{X}^\top)\bm{O}_l(\widehat{\bm{\theta}}) = \bm{C}_l - (\bm{I}_p \otimes \bm{X}^\top)\bm{O}_l(\bm{\theta}_0) + \ordR{0} 
        = O_p(1)
    \end{align*}
    element-wise, and hence
    \begin{align*}
        \bm{G}_{2,l}(\widehat{\bm{\theta}}) 
            &= \{\bm{C}_l - (\bm{I}_p \otimes \bm{X}^\top)\bm{O}_l(\widehat{\bm{\theta}})\}^\top \bm{Q}(\widehat{\bm{\theta}}) \{\bm{C}_l - (\bm{I}_p \otimes \bm{X}^\top)\bm{O}_l(\widehat{\bm{\theta}})\} \\
        &= O_p(1) O_p(n^{-1}) O_p(1) \\
        &= O_p(n^{-1}),
    \end{align*}
    element-wise by Assumption \ref{SM-regcond:X}.
    
    On combining all of the results above, we have 
    \begin{align*}
        \widehat{\bm{\Sigma}}_l^{-1/2}(\widehat{\bm{\theta}})\left\{\widehat{\bm{\mu}}_l(\widehat{\bm{\theta}}) - \bm{\mu}_l \right\} 
        &= \left\{\bm{G}_{1,l}(\widehat{\bm{\theta}}) + \bm{G}_{2,l} (\widehat{\bm{\theta}})\right\}^{-1/2} \left[\bm{C}_l^\top(\widehat{\bm{\beta}} - \bm{\beta}) +\bm{K}_l^\top \{\widehat{\bm{f}}(\widehat{\bm{\theta}}) - \bm{f}\} \right] \\
        &= \left\{\bm{G}_{1,l}(\widehat{\bm{\theta}}) + \ordR{-1}\right\}^{-1/2} \left[ \ordR{-1/2} +\bm{K}_l^\top \{\widehat{\bm{f}}(\widehat{\bm{\theta}}) - \bm{f}\} \right] \\
        &= \mathcal{Z}_d + \ordR{-1/2},
    \end{align*}
    and the required results follow.
\end{proof}

\subsection{Proof of Theorem 1}
\label{SM-sec:proof-thm1}

\begin{proof}
Let $M_n:=\max_{1\le l\le L}\|\bm T_{l,n}\|$ and $M_{0,n}:=\max_{1\le l\le L}\|\bm Z_l(\bm\theta_0)\|$, with distribution functions $F_n(t):=\mathbb P(M_n\le t)$ and $F_{0,n}(t):=\mathbb P(M_{0,n}\le t)$. Similarly, let $M_n^*:=\max_{1\le l\le L}\|\bm T_{l,n}^*\|$ and $M_{0,n}^*:=\max_{1\le l\le L}\|\bm Z_l^*(\widehat{\bm\theta})\|$, with conditional distribution functions $F_n^*(t):=\mathbb P^*(M_n^*\le t)$ and $F_{0,n}^*(t):=\mathbb P^*(M_{0,n}^*\le t)$, where $\mathbb P^*(\cdot)$ denotes probability under the bootstrap law conditional on the observed data $\bm y$. We want to show

\begin{align*}
    \sup_{t \in \mathbb{R}}|F_n^*(t)-F_n(t)| 
    & \le \sup_{t \in \mathbb{R}}|F_{0,n}(t)-F_n(t)| + \sup_{t \in \mathbb{R}}|F_{0,n}^*(t)-F_{0,n}(t)| + \sup_{t \in \mathbb{R}}|F_n^*(t)-F_{0,n}^*(t)| \\
    & \xrightarrow{p}0.
\end{align*}

We start by considering $\sup_{t\in\mathbb R}|F_{0,n}(t)-F_n(t)|$. By the reverse triangle inequality and Lemma~\ref{SM-lemma:order-reminder},
$$
\Delta_n:=|M_n-M_{0,n}| \le \max_{1\le l\le L}\|\bm T_{l,n}-\bm Z_l(\bm\theta_0)\| = O_p\left(\sqrt{\frac{\log L}{n}}\right).
$$
For any deterministic $\epsilon_n>0$, let $E_n=\{\Delta_n\le\epsilon_n\}$. On $E_n$, we have $|M_n-M_{0,n}|\le\epsilon_n$, and hence $M_{0,n}-\epsilon_n\le M_n\le M_{0,n}+\epsilon_n$. Therefore, $\{M_{0,n}\le t-\epsilon_n\}\cap E_n\subseteq\{M_n\le t\}.$
For the upper inclusion, if $M_n\le t$, then either $E_n$ occurs, in which case $M_{0,n}\le M_n+\epsilon_n\le t+\epsilon_n$, or $E_n^c$ occurs. Hence, $\{M_n\le t\}\subseteq\{M_{0,n}\le t+\epsilon_n\}\cup E_n^c.$
Thus,
$$
\{M_{0,n}\le t-\epsilon_n\}\cap\{\Delta_n\le\epsilon_n\} \subseteq
\{M_n\le t\} \subseteq \{M_{0,n}\le t+\epsilon_n\}\cup\{\Delta_n>\epsilon_n\}.
$$
Taking probabilities then gives
$$
F_{0,n}(t-\epsilon_n)-\mathbb P(\Delta_n>\epsilon_n)
\le F_n(t)
\le F_{0,n}(t+\epsilon_n)+\mathbb P(\Delta_n>\epsilon_n).
$$
Subtracting $F_{0,n}(t)$ bounds the difference by the increments
$F_{0,n}(t+\epsilon_n)-F_{0,n}(t)$ and $F_{0,n}(t)-F_{0,n}(t-\epsilon_n)$.
Each is at most $\mathbb P(|M_{0,n}-t|\le\epsilon_n)$. Thus
$$
\sup_{t\in\mathbb R}|F_{0,n}(t)-F_n(t)| \le \sup_{x\in\mathbb R}\mathbb P(|M_{0,n}-x|\le\epsilon_n) + \mathbb P(\Delta_n>\epsilon_n).
$$
The small-interval probability is expressed using the Gaussian oracle $M_{0,n}$ so that Gaussian anti-concentration applies. To establish $\sup_{t\in\mathbb R}|F_{0,n}(t)-F_n(t)|\to0$, it suffices to choose $\epsilon_n$ such that the two terms on the right-hand side vanish simultaneously. We first consider the anti-concentration term. Let $\mathcal T_L=[L]\times\mathbb S^{d-1}$ and define $X_{l,\bm v}:=\bm v^\top\bm Z_l(\bm\theta_0)$ for $(l,\bm v)\in\mathcal T_L$, where $\|\bm v\|=1$. Since $\bm Z_l(\bm\theta_0)\sim N_d(\bm0,\bm I_d)$, we have $\mathbb E(X_{l,\bm v})=0$ and $\mathbb E(X_{l,\bm v}^2)=1$, while
$$
M_{0,n}=\max_{1\le l\le L}\|\bm Z_l(\bm\theta_0)\| = \sup_{(l,\bm v)\in\mathcal T_L}\bm v^\top\bm Z_l(\bm\theta_0) = \sup_{(l,\bm v)\in\mathcal T_L}X_{l,\bm v}.
$$
The process is separable since $\mathbb S^{d-1}$ is separable and $\bm v\mapsto\bm v^\top\bm Z_l(\bm\theta_0)$ is continuous. Hence, by Lemmas~\ref{SM-lemma:order-leader} and \ref{SM-lemma:thm2.1-CCK},
$$
\sup_{x\in\mathbb R}\mathbb P(|M_{0,n}-x|\le\epsilon_n) \le 4\epsilon_n\{ \mathbb E(M_{0,n})+1\} \le C\epsilon_n\sqrt{\log L}.
$$
Thus, for the first term to vanish, we require $\epsilon_n\sqrt{\log L}\to 0$, or equivalently $\epsilon_n = o\{(\log L)^{-1/2}\}$.
For the second term, Lemma~\ref{SM-lemma:order-reminder} gives $\Delta_n = O_p\{\sqrt{\log L/n}\}$. Therefore, to ensure $\mathbb P(\Delta_n>\epsilon_n) \to 0$, it is sufficient that $\sqrt{\frac{\log L}{n}} = o(\epsilon_n).$
Hence, $\epsilon_n$ should be chosen between the two rates,
$$
\sqrt{\frac{\log L}{n}} \ll \epsilon_n \ll \frac{1}{\sqrt{\log L}}.
$$
Such a sequence exists precisely under $\log L=o(\sqrt n)$. For example, take $\epsilon_n = n^{-1/4}$. Then
$$
\epsilon_n\sqrt{\log L} = \left(\frac{\log L}{\sqrt n}\right)^{1/2}\to 0, \qquad \frac{\sqrt{\log L/n}}{\epsilon_n} = \left(\frac{\log L}{\sqrt n}\right)^{1/2} \to 0.
$$
Consequently, both terms vanish and therefore
\begin{align}
    \sup_{t\in\mathbb R}|F_{0,n}(t)-F_n(t)| \to 0.
    \label{SM-eq:orgin-cdf-converge}
\end{align}

We next consider $\sup_{t\in\mathbb R}|F_{0,n}^*(t)-F_{0,n}(t)|$. Let
$$
\bm W(\bm\theta):=\{\bm I_m+\bm\Lambda^\top\bm\Psi^{-1}\bm\Lambda\}^{-1},
\qquad
\bm A_l(\bm\theta):=\bm G_{1,l}^{-1/2}(\bm\theta)\bm K_l^\top\{\bm W^{1/2}(\bm\theta)\otimes\bm I_n\}.
$$
Introduce an auxiliary random vector $\bm g\sim N_{mn}(\bm0,\bm I_{mn})$, independent of the observed data, and define
$$
\bm Z_l^g(\bm\theta):=\bm A_l(\bm\theta)\bm g, \qquad M_{0,n}^g(\bm\theta):=\max_{1\le l\le L}\|\bm Z_l^g(\bm\theta)\|.
$$
We first verify that these auxiliary Gaussian vectors reproduce the joint laws of the corresponding oracle quantities. For any $r,s\in[L]$,
\begin{align*}
\operatorname{Cov}_g\{\bm Z_r^g(\bm\theta_0),\bm Z_s^g(\bm\theta_0)\}
&=\bm A_r(\bm\theta_0)\bm A_s^\top(\bm\theta_0) \\
&=\bm G_{1,r}^{-1/2}(\bm\theta_0)\bm K_r^\top
\{\bm W(\bm\theta_0)\otimes\bm I_n\}\bm K_s
\bm G_{1,s}^{-1/2}(\bm\theta_0) \\
&=\operatorname{Cov}\{\bm Z_r(\bm\theta_0),\bm Z_s(\bm\theta_0)\}.
\end{align*}
Both collections are centered jointly Gaussian, and hence $\{\bm Z_l^g(\bm\theta_0)\}_{l=1}^L \overset{d}{=} \{\bm Z_l(\bm\theta_0)\}_{l=1}^L.$
Similarly, conditional on the observed data, the bootstrap oracle prediction error has covariance $\bm W(\widehat{\bm\theta})\otimes\bm I_n$. Therefore, $\{\bm Z_l^g(\widehat{\bm\theta})\}_{l=1}^L \overset{d^*}{=} \{\bm Z_l^*(\widehat{\bm\theta})\}_{l=1}^L.$
Consequently, $M_{0,n}^g(\bm\theta_0)\overset{d}{=}M_{0,n}$ and, conditional on the data, $M_{0,n}^g(\widehat{\bm\theta})\overset{d^*}{=}M_{0,n}^*$.

We now compare the two coupled Gaussian maxima under the same realisation of $\bm g$. By the reverse triangle inequality,
$$
|M_{0,n}^g(\widehat{\bm\theta})-M_{0,n}^g(\bm\theta_0)|
\le\max_{1\le l\le L}\|\{\bm A_l(\widehat{\bm\theta})-\bm A_l(\bm\theta_0)\}\bm g\|.
$$
To determine the order of the right-hand side, note that
\begin{align*}
\bm A_l(\widehat{\bm\theta})-\bm A_l(\bm\theta_0) = {}
&\{\bm G_{1,l}^{-1/2}(\widehat{\bm\theta})-\bm G_{1,l}^{-1/2}(\bm\theta_0)\} \bm K_l^\top\{\bm W^{1/2}(\widehat{\bm\theta})\otimes\bm I_n\} \\
& \quad + \bm G_{1,l}^{-1/2}(\bm\theta_0)\bm K_l^\top \{[\bm W^{1/2}(\widehat{\bm\theta})-\bm W^{1/2}(\bm\theta_0)]\otimes\bm I_n\}.
\end{align*}
By Lemma~\ref{SM-lemma:order_of_G1andG2}, the uniform lower eigenvalue bound on $\bm G_{1,l}(\bm\theta_0)$, and local Lipschitz continuity of the inverse square-root map imply $\max_{1\le l\le L} \|\bm G_{1,l}^{-1/2}(\widehat{\bm\theta}) - \bm G_{1,l}^{-1/2}(\bm\theta_0)\| = O_p(n^{-1/2}).$
Moreover, smoothness of $\bm W^{1/2}(\bm\theta)$ and $\|\widehat{\bm\theta}-\bm\theta_0\| = O_p(n^{-1/2})$ imply $\|\bm W^{1/2}(\widehat{\bm\theta}) - \bm W^{1/2}(\bm\theta_0)\| = O_p(n^{-1/2}).$
Together with the uniform boundedness of $\bm K_l$, $\bm G_{1,l}^{-1/2}(\bm\theta_0)$ and $\bm W^{1/2}(\widehat{\bm\theta})$ in Assumptions \ref{SM-regcond:true-para}, \ref{SM-regcond:designed-mat}, and \ref{SM-regcond:PMSE}, it follows that
$$
\max_{1\le l\le L}\|\bm A_l(\widehat{\bm\theta})-\bm A_l(\bm\theta_0)\| = O_p(n^{-1/2}).
$$
Conditional on the observed data, $\bm Z_l^g(\widehat{\bm\theta})-\bm Z_l^g(\bm\theta_0) = \{\bm A_l(\widehat{\bm\theta})-\bm A_l(\bm\theta_0)\} \bm g$ is a centred Gaussian vector satisfying
$$
\max_{1\le l\le L} \zeta_{\max}\left[ \operatorname{Var}_g\{\bm Z_l^g(\widehat{\bm\theta})-\bm Z_l^g(\bm\theta_0)\mid\bm y\} \right] \le \max_{1\le l\le L}\|\bm A_l(\widehat{\bm\theta})-\bm A_l(\bm\theta_0)\|^2 = O_p(n^{-1}).
$$
Since $d$ is fixed, a Gaussian tail bound (similar to the proof of Lemma \ref{SM-lemma:order-leader} in Section \ref{SM-sec:supplement-proof-of-lemma}) together with the union bound gives
$$
\max_{1\le l\le L} \|\bm Z_l^g(\widehat{\bm\theta}) - \bm Z_l^g(\bm\theta_0)\| = O_{P_g}\left(\sqrt{\frac{\log L}{n}}\right) \quad\text{in probability},
$$
where $\mathbb P_g(\cdot\mid\bm y)$ is the conditional law of $\bm g\sim N_{mn}(\bm0,\bm I_{mn})$, with the data held fixed. We write $R_n=O_{P_g}(a_n)$ in probability when, for every $\epsilon>0$, there exists an constant $0<M<\infty$, such that
$$
\limsup_{n\to\infty} \mathbb P\{\mathbb P_g(|R_n|>Ma_n\mid\bm y)>\epsilon\}=0.
$$
Consequently,
$$
|M_{0,n}^g(\widehat{\bm\theta})-M_{0,n}^g(\bm\theta_0)| \le \max_{1\le l\le L} \|\bm Z_l^g(\widehat{\bm\theta}) - \bm Z_l^g(\bm\theta_0)\| = O_{P_g}\left(\sqrt{\frac{\log L}{n}}\right)
$$
in probability.

We are now in the same setting as for the first difference. In particular, for any deterministic $\epsilon_n>0$, the same probability-sandwich argument gives
$$
\sup_{t\in\mathbb R}|F_{0,n}^*(t)-F_{0,n}(t)| \le \sup_{x\in\mathbb R}\mathbb P(|M_{0,n}-x|\le\epsilon_n) + \mathbb P_g\left\{|M_{0,n}^g(\widehat{\bm\theta})-M_{0,n}^g(\bm\theta_0)|>\epsilon_n \,\middle|\,\bm y \right\}.
$$
By Lemmas~\ref{SM-lemma:order-leader} and \ref{SM-lemma:thm2.1-CCK}, the first term is bounded by $C\epsilon_n\sqrt{\log L}$. For the second term, the preceding result gives $|M_{0,n}^g(\widehat{\bm\theta}) -M_{0,n}^g(\bm\theta_0)| = O_{P_g}\left(\sqrt{\frac{\log L}{n}}\right) \quad\text{in probability}.$
Hence, taking again $\epsilon_n=n^{-1/4}$, the condition $\log L=o(\sqrt n)$ ensures $\epsilon_n\sqrt{\log L} \to 0$ and $\sqrt{\frac{\log L}{n}}=o(\epsilon_n),$
so that both terms on the right-hand side converge to zero, with the second converging in probability under the original sampling law. Therefore,
\begin{align}
    \sup_{t\in\mathbb R}|F_{0,n}^*(t)-F_{0,n}(t)| \xrightarrow{p} 0.
    \label{SM-eq:oracle-cdf-coverge}
\end{align}

Finally, consider $\sup_{t\in\mathbb R}|F_n^*(t)-F_{0,n}^*(t)|$. Conditional on the observed data, the bootstrap sample is generated under $(\widehat{\bm\beta},\widehat{\bm\theta})$. Define the bootstrap oracle predictor $\check{\bm f}^{\,*}(\widehat{\bm\theta}) :=\{(\widehat{\bm\Lambda}^{\top} \widehat{\bm V}^{-1}) \otimes \bm I_n\}\{\bm y^* - (\bm I_p\otimes\bm X)\widehat{\bm\beta}\},$ and $\bm Z_l^*(\widehat{\bm\theta}) := \bm G_{1,l}^{-1/2}(\widehat{\bm\theta})\bm K_l^\top \{\check{\bm f}^{\,*}(\widehat{\bm\theta})-\bm f^*\}.$
By the same Gaussian prediction-error calculation as for the original oracle,
$\bm Z_l^*(\widehat{\bm\theta})\mid\bm y\sim N_d(\bm0,\bm I_d)$.
Under the regularity conditions \ref{SM-regcond:X} and \ref{SM-regcond:MLE}, the parametric bootstrap estimators satisfy
$$
\|\widehat{\bm\theta}^{\,*} - \widehat{\bm\theta}\| = O_{p^*}(n^{-1/2}), \qquad \|\widehat{\bm\beta}^{\,*} - \widehat{\bm\beta}\| = O_{p^*}(n^{-1/2})
$$
in probability. Hence, conditionally on the observed data, the argument of Lemma~\ref{SM-lemma:order-reminder} applies with $(\bm\theta_0,\bm\beta_0)$ replaced by $(\widehat{\bm\theta},\widehat{\bm\beta})$, and consequently $
\max_{1\le l\le L} \|\bm T_{l,n}^* - \bm Z_l^*(\widehat{\bm\theta})\| = O_{p^*}\left(\sqrt{\frac{\log L}{n}}\right)$
in probability.
It follows from the reverse triangle inequality that
$$
|M_n^*-M_{0,n}^*| \le \max_{1\le l\le L} \|\bm T_{l,n}^*-\bm Z_l^*(\widehat{\bm\theta})\| = O_{p^*}\left(\sqrt{\frac{\log L}{n}}\right)
$$
in probability. We are therefore again in the same setting as for the first difference. Conditional on the observed data, $\{\bm Z_l^*(\widehat{\bm\theta})\}_{l=1}^L$ is a centred standardised Gaussian process, so the Lemma~\ref{SM-lemma:thm2.1-CCK} and the same probability-sandwich argument yield, for any deterministic $\epsilon_n>0$,
$$
\sup_{t\in\mathbb R}|F_n^*(t) - F_{0,n}^*(t)| \le \sup_{x\in\mathbb R}\mathbb P^*(|M_{0,n}^* - x|\le\epsilon_n) + \mathbb P^*(|M_n^*-M_{0,n}^*| > \epsilon_n).
$$
Taking again $\epsilon_n = n^{-1/4}$, the condition $\log L = o(\sqrt n)$ makes both terms converge to zero in probability. Consequently,
\begin{align}
    \sup_{t\in\mathbb R}|F_n^*(t)-F_{0,n}^*(t)| \xrightarrow{p} 0.
    \label{SM-eq:bootstrap-cdf-coverge}
\end{align}
Combining Equations \eqref{SM-eq:orgin-cdf-converge}, \eqref{SM-eq:oracle-cdf-coverge}, and \eqref{SM-eq:bootstrap-cdf-coverge}, we have 
\begin{align*}
    \sup_{t \in \mathbb{R}}|F_n^*(t)-F_n(t)| \xrightarrow{p} 0,
\end{align*}
as we require to prove the first part of the theorem.

To deduce coverage, we bracket the random bootstrap critical value by deterministic oracle quantiles. By the triangle inequality, $\|F_n^* - F_{0,n}\|_\infty \xrightarrow{p} 0$. Let $\eta = 1-\alpha$, $q_{0,n}(u) = F_{0,n}^{-1}(u)$, and $\widehat c_n = (F_n^*)^{-1}(\eta)$, where $F^{-1}(u)=\inf\{t:F(t)\ge u\}$. We use $\widehat{\mathcal S}_{1-\alpha,l}^{\mathrm{BS}_0}$ to denote the SPR constructed using this ideal bootstrap critical value. Since each $\|\bm Z_l(\bm\theta_0)\|$ has a continuous distribution,
$$
\mathbb P(M_{0,n}=t) \le \sum_{l=1}^L \mathbb P(\|\bm Z_l(\bm\theta_0)\|=t) = 0 \quad \forall t \ge 0,
$$
so $F_{0,n}$ is continuous. For any fixed $0<\delta<\min(\eta,1-\eta)$, define $E_{n,\delta} = \{\|F_n^* - F_{0,n}\|_\infty \le \delta\}$. On \(E_{n,\delta}\), the bootstrap and oracle distribution functions satisfy
$$
F_{0,n}(t)-\delta \le F_n^*(t) \le F_{0,n}(t)+\delta,
\quad \forall t\in\mathbb R.
$$
For any \(t<q_{0,n}(\eta-\delta)\), the definition of $q_{0,n}(u)$
gives \(F_{0,n}(t)<\eta-\delta\), so \(F_n^*(t)<\eta\). At the upper
endpoint \(t=q_{0,n}(\eta+\delta)\), we have \(F_{0,n}(t)\ge\eta+\delta\),
so \(F_n^*(t)\ge\eta\). Since
\(\widehat c_n=\inf\{t\in\mathbb R : F_n^*(t)\ge\eta\}\), it follows that
\begin{equation}\label{SM-eq:quantile-bracket}
    q_{0,n}(\eta-\delta)
    \le \widehat c_n
    \le q_{0,n}(\eta+\delta).
\end{equation}
Consequently,
$$
\{M_n \le q_{0,n}(\eta-\delta)\}\cap E_{n,\delta} \subseteq \{M_n \le \widehat c_n\} \subseteq \{M_n \le q_{0,n}(\eta+\delta)\}\cup E_{n,\delta}^c.
$$
Taking probabilities and using the continuity of $F_{0,n}$ gives
$$
\eta-\delta-r_n-\mathbb P(E_{n,\delta}^c) \le \mathbb P(M_n \le \widehat c_n) \le \eta+\delta+r_n+\mathbb P(E_{n,\delta}^c),
$$
where $r_n = \|F_n-F_{0,n}\|_\infty = o(1)$. Since $\mathbb P(E_{n,\delta}^c)\to0$, letting $n\to\infty$ and then $\delta\downarrow0$ yields
$$
\mathbb P(M_n \le \widehat c_n) = \mathbb P\left(\bigcap_{l=1}^L\left\{\bm\mu_l\in\widehat{\mathcal S}_{1-\alpha,l}^{\mathrm{BS}_0}\right\}\right) \longrightarrow 1-\alpha.
$$

It remains to account for the use of an empirical bootstrap CDF (and hence the bootstrap quantile) in place of $\widehat{c}_n$.

Conditional on the observed data $\bm y$, let $M_n^{*(1)},\ldots,M_n^{*(B)}$ be independent bootstrap maxima with CDF $F_n^*$, and define
$$
\widehat F_{n,B}^*(t) = \frac{1}{B}\sum_{b=1}^{B}\mathbf{1}\{M_n^{*(b)}\le t\}.
$$
By the Dvoretzky--Kiefer--Wolfowitz--Massart (DKWM) inequality \citep{massart1990tight}, for every $\varepsilon>0$,
$$
\mathbb P^*\left(\|\widehat F_{n,B}^*-F_n^*\|_\infty>\varepsilon\right) \le 2\exp(-2B\varepsilon^2).
$$
Thus, $\|\widehat F_{n,B}^*-F_n^*\|_\infty = O_{p^*}(B^{-1/2})$. If $B\to\infty$, the triangle inequality gives
$$
\|\widehat F_{n,B}^*-F_{0,n}\|_\infty \le \|\widehat F_{n,B}^*-F_n^*\|_\infty + \|F_n^*-F_{0,n}\|_\infty \xrightarrow{p}0.
$$
Let \(B\to\infty\), and fix \(0<\delta<\min(\eta,1-\eta)\).
Define $\widehat E_{n,\delta} =
\bigl\{\|\widehat F_{n,B}^*-F_{0,n}\|_\infty\le\delta\bigr\}.$
By the triangle inequality and the DKWM inequality,
\[
\mathbb P(\widehat E_{n,\delta}^{\,c})
\le
2\exp(-B\delta^2/2)
+
\mathbb P\!\left(\|F_n^*-F_{0,n}\|_\infty>\delta/2\right)
\rightarrow 0.
\]
The first term vanishes because \(B\to\infty\), and the second vanishes
by the uniform convergence in probability established above.

On \(\widehat E_{n,\delta}\), the empirical bootstrap CDF is below
\(\eta\) for every \(t<q_{0,n}(\eta-\delta)\) and is at least \(\eta\)
at \(t=q_{0,n}(\eta+\delta)\). Thus, as in \eqref{SM-eq:quantile-bracket},
the empirical bootstrap critical value satisfies
\[
q_{0,n}(\eta-\delta)
\le \widehat c_{n,B}
\le q_{0,n}(\eta+\delta),
\qquad
\widehat c_{n,B}=(\widehat F_{n,B}^*)^{-1}(\eta).
\]
Bounding the coverage probability using these two deterministic quantiles
and recalling that \(r_n=\|F_n-F_{0,n}\|_\infty\), we obtain
\[
\eta-\delta-r_n-\mathbb P(\widehat E_{n,\delta}^{\,c})
\le \mathbb P(M_n\le\widehat c_{n,B})
\le \eta+\delta+r_n+\mathbb P(\widehat E_{n,\delta}^{\,c}).
\]
Since \(r_n\to0\) and \(\mathbb P(\widehat E_{n,\delta}^{\,c})\to0\)
for each fixed \(\delta\), letting \(n\to\infty\) and then
\(\delta\downarrow0\) gives
\[
\mathbb P\!\left(
  \bigcap_{l=1}^L
  \left\{\bm\mu_l\in\widehat{\mathcal S}_{1-\alpha,l}^{\mathrm{BS}}\right\}
\right)
=
\mathbb P(M_n\le\widehat c_{n,B})
\rightarrow \eta=1-\alpha,
\]
where \(\widehat{\mathcal S}_{1-\alpha,l}^{\mathrm{BS}}\) is the prediction
region for target \(l\) constructed using \(\widehat c_{n,B}\).
This completes the proof.

\end{proof}

\subsection{Proof of Proposition 1}
\label{SM-sec:proof-prop2}
\setcounter{equation}{0}

The normality of $\bm\xi - \tilde{\bm\xi}$ is obvious, as $\tilde{\bm\xi}$ is a linear function of $\boldsymbol{y}$ according to the mixed model equation for an FA model with $\boldsymbol{y}$ and $\bm\xi$ themselves being normal.

Next, by the properties of BLUP, we have
\begin{equation*}
    \mathbb{E}[(\bm\xi - \tilde{\bm\xi}) \mid \boldsymbol{y}] = \mathbb{E}[\bm\xi \mid \boldsymbol{y}] - \mathbb{E}\{\mathbb{E}[\bm\xi \mid \boldsymbol{y}]\} = \mathbf{0}.
\end{equation*}
Therefore, the marginal mean satisfies.
$$
\mathbb{E}[\bm\xi - \tilde{\bm\xi}] = \mathbb{E}\{\mathbb{E}[(\bm\xi - \tilde{\bm\xi}) \mid \boldsymbol{y}]\} = \mathbf{0}.
$$
For the variance, denote $\bm{\Omega}_\xi:=\bm{P} + \bm{A}^\top\bm{W}^{-1}\bm{A}$. Then using the mixed model equations again, we can rewrite $\bm\xi - \tilde{\bm\xi}$ as 
\begin{align*}
    \bm\xi - \tilde{\bm\xi} &= \bm\xi - \bm{\Omega}_\xi^{-1}\bm{A}^\top \bm{W}^{-1}\boldsymbol{y} \\
    &= \bm\xi - \bm{\Omega}_\xi^{-1}\bm{A}^\top \bm{W}^{-1}(\bm{A}\bm\xi+\boldsymbol{\epsilon}) \\
    &= (\bm{I} - \bm{\Omega}_\xi^{-1}\bm{A}^\top \bm{W}^{-1}\bm{A})\bm\xi - \bm{\Omega}_\xi^{-1}\bm{A}^\top \bm{W}^{-1}\boldsymbol{\epsilon} \\ 
    &= \bm{\Omega}_\xi^{-1}\bm{P}\bm\xi - \bm{\Omega}_\xi^{-1}\bm{A}^\top \bm{W}^{-1}\boldsymbol{\epsilon}.
\end{align*}
It follows that 
\begin{align*}
    \mathrm{Var}(\bm\xi - \tilde{\bm\xi}) &= \bm{\Omega}_\xi^{-1}\bm{P}\mathrm{Var}(\bm\xi)\bm{P}^\top\bm{\Omega}_\xi^{-1} + \bm{\Omega}_\xi^{-1}\bm{A}^\top\bm{W}^{-1}\mathrm{Var}(\boldsymbol{\epsilon})\bm{W}^{-1}\bm{A}\mathbf{Q}^{-1} \\
    &= \bm{\Omega}_\xi^{-1}\bm{P}\bm{\Omega}_\xi^{-1} + \bm{\Omega}_\xi^{-1}\bm{A}^\top\bm{W}^{-1}\bm{W}\bm{W}^{-1}\bm{A}\bm{\Omega}_\xi^{-1} \\
    &= \bm{\Omega}_\xi^{-1}(\bm{P} + \bm{A}^\top\bm{W}^{-1}\bm{A})\bm{\Omega}_\xi^{-1} \\
    &= \bm{\Omega}_\xi^{-1} = (\bm{P} + \bm{A}^\top\bm{W}^{-1}\bm{A})^{-1},
\end{align*}
as required.

\subsection{Proofs of lemmas}
\setcounter{equation}{0} \label{SM-sec:supplement-proof-of-lemma}

\begin{proof}[Proof of Lemma~\ref{SM-lemma:variance-component-approx}]
    Write 
    \begin{align*}
    \widehat{\bm{V}} - \bm{V}_0 &=  \widehat{\bm{\Lambda}}\widehat{\bm{\Lambda}}^\top + \widehat{\bm{\Psi}} - (\bm{\Lambda}_0\bm{\Lambda}_0^\top + \bm{\Psi}_0) \\
    &= (\widehat{\bm{\Lambda}} - \bm{\Lambda}_0) \bm{\Lambda}_0^\top + \bm{\Lambda}_0 (\widehat{\bm{\Lambda}} - \bm{\Lambda}_0)^\top + (\widehat{\bm{\Lambda}} - \bm{\Lambda}_0)(\widehat{\bm{\Lambda}} - \bm{\Lambda}_0)^\top + (\widehat{\bm{\Psi}} - \bm{\Psi}_0) \\
    &= \bm{A}(\widehat{\bm{\theta}}, \bm{\theta}_0) + (\widehat{\bm{\Lambda}} - \bm{\Lambda}_0)(\widehat{\bm{\Lambda}} - \bm{\Lambda}_0)^\top. 
    \end{align*}
    Since $\|\widehat{\bm{\Lambda}} - \bm{\Lambda}_0\|_F = O(n^{-1/2})$, then $\|(\widehat{\bm{\Lambda}} - \bm{\Lambda}_0)(\widehat{\bm{\Lambda}} - \bm{\Lambda}_0)^\top\|_F = O_p(n^{-1})$, and hence $\widehat{\bm{V}} = \bm{V}_0 + \bm{A}(\widehat{\bm{\theta}}, \bm{\theta}_0) + \ordR{-1}$. Furthermore, the three submatrices comprising $\bm{A}(\widehat{\bm{\theta}}, \bm{\theta}_0)$ are, and hence $\bm{A}(\widehat{\bm{\theta}}, \bm{\theta}_0)$ itself is, $O_p(n^{-1/2})$ element-wise. It follows that by a series expansion, we obtain
    \begin{align*}
    \widehat{\bm{V}}^{-1} &= [\bm{V}_0 + \{\bm{A}(\widehat{\bm{\theta}}, \bm{\theta}_0) + \ordR{-1}\}]^{-1} \\
    &= \bm{V}^{-1}_0 - \bm{V}^{-1}_0 \{\bm{A}(\widehat{\bm{\theta}}, \bm{\theta}_0) + \ordR{-1}\} \bm{V}^{-1}_0 + \ordR{-1} \\
    &= \bm{V}^{-1}_0 - \bm{V}^{-1}_0 \bm{A}(\widehat{\bm{\theta}}, \bm{\theta}_0)\bm{V}^{-1}_0 + \ordR{-1}.
    \end{align*}
    The required result follows from this.
\end{proof}

\begin{proof}[Proof of Lemma~\ref{SM-lemma:order_of_G1andG2}]
    For the first part, we have
    \begin{align*}
    \max_{1\le l\le L}\|\bm G_{1,l}(\widehat{\bm\theta})-\bm G_{1,l}(\bm\theta_0)\| 
    &\le \max_{1\le l\le L}\|\bm K_l\|^2 \bigl\| \left\{\bm I_{mn} +(\widehat{\bm\Lambda}^{\top}\widehat{\bm\Psi}^{-1}\widehat{\bm\Lambda})\otimes\bm I_n\right\}^{-1} \\
    & - \left\{\bm I_{mn}+(\bm\Lambda_0^\top\bm\Psi_0^{-1}\bm\Lambda_0)\otimes\bm I_n\right\}^{-1}
    \bigr\|.
    \end{align*}
    
    By Assumption \ref{SM-regcond:designed-mat}, $\max_{1\le l\le L}\|\bm K_l\| = O(1)$. Moreover, since the map $\bm\theta\mapsto \left\{\bm I_m+\bm\Lambda^\top\bm\Psi^{-1}\bm\Lambda\right\}^{-1}$ is smooth in a neighbourhood of $\bm\theta_0$, and $\|\widehat{\bm\theta}-\bm\theta_0\| = O_p(n^{-1/2})$, then we have
    $$
    \left\|\{\bm I_m + \widehat{\bm\Lambda}^{\top}\widehat{\bm\Psi}^{-1}\widehat{\bm\Lambda}\}^{-1} - \left\{\bm I_m + \bm\Lambda_0^\top\bm\Psi_0^{-1}\bm\Lambda_0\right\}^{-1} \right\| = O_p(n^{-1/2}).
    $$
    Since $\bm I_{mn} + (\bm\Lambda^\top\bm\Psi^{-1}\bm\Lambda) \otimes \bm I_n = \left(\bm I_m + \bm\Lambda^\top\bm\Psi^{-1}\bm\Lambda\right) \otimes \bm I_n,$
    and taking the Kronecker product with $\bm I_n$ does not change the operator norm, it follows that
    $$
    \left\| \{\bm I_{mn}+(\widehat{\bm\Lambda}^{\top}\widehat{\bm\Psi}^{-1}\widehat{\bm\Lambda})\otimes\bm I_n\}^{-1} - \left\{\bm I_{mn} + (\bm\Lambda_0^\top\bm\Psi_0^{-1}\bm\Lambda_0) \otimes \bm I_n\right\}^{-1} \right\| = O_p(n^{-1/2}).
    $$
    Hence,
    $$
    \max_{1\le l\le L}\|\bm G_{1,l}(\widehat{\bm\theta})-\bm G_{1,l}(\bm\theta_0)\| = O_p(n^{-1/2}),
    $$
    as required.

\noindent Turning to the second part, let $\bm D_l(\bm\theta) = \bm C_l-(\bm I_p\otimes\bm X)^\top\bm O_l(\bm\theta)$. We first show that
    $$
    \max_{1\le l\le L}\left\| \bm D_l(\bm\theta_0)\right\| = O(1).
    $$
    Recall that $\bm O_l(\bm\theta_0) = \{(\bm V_0^{-1} \bm\Lambda_0) \otimes \bm I_n\} \bm K_l$. The block definition in Assumption~\ref{SM-regcond:designed-mat} gives the exact identity
    $$
    (\bm I_p\otimes\bm X)^\top\bm O_l(\bm\theta_0) = \sum_{i\in\mathcal S_l}\{(\bm V_0^{-1}\bm\Lambda_0)\otimes\bm x_i\}\bm K_{li}^\top.
    $$
    Consequently,
    $$
    \left\|(\bm I_p\otimes\bm X)^\top\bm O_l(\bm\theta_0)\right\|
    \le \sum_{i\in\mathcal S_l}\|\bm x_i\|\|\bm V_0^{-1}\bm\Lambda_0\|\|\bm K_{li}\|.
    $$
    By Assumptions \ref{SM-regcond:X}, \ref{SM-regcond:designed-mat} and \ref{SM-regcond:PMSE}, we have $\max_i\|\bm x_i\| = O(1)$, $\sup_l|\mathcal S_l|\le S_K<\infty$, $\sup_{l,i}\|\bm K_{li}\|\le C_K<\infty$, and $\|\bm V_0^{-1}\bm\Lambda_0\| = O(1)$. Hence,
    $$
    \max_{1\le l\le L}\left\|(\bm I_p\otimes\bm X)^\top\bm O_l(\bm\theta_0)\right\| = O(1).
    $$
    Since Assumption \ref{SM-regcond:designed-mat} also gives $\max_{1\le l\le L}\|\bm C_l\| = O(1)$, it follows that
    $$
    \max_{1\le l\le L}\left\|\bm D_{l}(\bm\theta _0)\right\| \le \max_{1\le l\le L}\|\bm C_l\|+\max_{1\le l\le L}\left\|(\bm I_p\otimes\bm X)^\top\bm O_l(\bm\theta_0)\right\| = O(1).
    $$
    
    Now we show $\max_{1\le l\le L}\|\bm D_{l}(\widehat{\bm\theta})\| = O_p(1)$. We have
    $$
    \bm D_l(\widehat{\bm\theta})-\bm D_l(\bm\theta_0) = -(\bm I_p\otimes\bm X)^\top\{\bm O_l(\widehat{\bm\theta})-\bm O_l(\bm\theta_0)\}.
    $$
    To bound this difference, take a convex neighbourhood $\delta(\bm\theta_0)$ of $\bm\theta_0$ whose closure lies in $\omega$. With probability tending to one, $\widehat{\bm\theta}\in\mathcal \delta(\bm\theta_0)$. Integrating the derivative along the line segment from $\bm\theta_0$ to $\widehat{\bm\theta}$ and applying Assumption~\ref{SM-regcond:D-mat} gives
    $$
    \begin{aligned}
    \max_{1\le l\le L}\|\bm D_l(\widehat{\bm\theta})-\bm D_l(\bm\theta_0)\| &= \max_{1\le l\le L}\|(\bm I_p\otimes\bm X)^\top\{\bm O_l(\widehat{\bm\theta})-\bm O_l(\bm\theta_0)\}\| \\
    &\le \max_{1\le l\le L}\sum_{j=1}^{h}
    \sup_{\bm\theta\in\delta(\bm\theta_0)} \left\| \frac{\partial}{\partial\theta_j}(\bm I_p\otimes\bm X)^\top\bm O_l(\bm\theta)
    \right\| |\widehat\theta_j-\theta_{0,j}| \\
    &\le C\sum_{j=1}^{h}|\widehat\theta_j-\theta_{0,j}|
    \le C \sqrt{h}\|\widehat{\bm\theta}-\bm\theta_0\| = O_p(n^{-1/2}),
    \end{aligned}
    $$
    for a sufficiently large constant $C > 0$, where $h=pm-m(m-1)/2+p$ was defined in Assumption~\ref{SM-regcond:true-para}, and the last inequality is Cauchy--Schwarz.. Therefore,
    $$
    \max_{1\le l\le L}\|\bm D_l(\widehat{\bm\theta})\| \le \max_{1\le l\le L}\|\bm D_l(\bm\theta_0)\| + \max_{1\le l\le L}\|\bm D_l(\widehat{\bm\theta}) - \bm D_l(\bm\theta_0)\| = O(1) + O_p(n^{-1/2}) = O_p(1),
    $$
    and hence,
    $$
    \max_{1\le l\le L}\left\|\bm D_l(\widehat{\bm\theta})\right\| = O_p(1).
    $$

    Finally, by Assumption \ref{SM-regcond:PMSE} and $\|\widehat{\bm\theta}-\bm\theta_0\| = O_p(n^{-1/2})$ we have $\|\widehat{\bm V}\| = O_p(1)$, while Assumption \ref{SM-regcond:X} gives $\|(\bm X^\top\bm X)^{-1}\| = O(n^{-1})$. Hence,
    $$
    \|\bm Q(\widehat{\bm\theta})\| = \|\widehat{\bm V}\otimes(\bm X^\top\bm X)^{-1}\| = \|\widehat{\bm V}\|\|(\bm X^\top\bm X)^{-1}\| = O_p(n^{-1}),
    $$
    and we obtain,
    $$
    \begin{aligned}
    \max_{1\le l\le L}\|\bm G_{2,l}(\widehat{\bm\theta})\| &\le \left\{\max_{1\le l\le L} \left\|\bm D_l(\widehat{\bm\theta})\right\|^2 \right\} \|\bm Q(\widehat{\bm\theta})\| \\
    &= O_p(1)O_p(n^{-1}) = O_p(n^{-1}),
    \end{aligned}
    $$
    as required.
\end{proof}

\begin{proof}[Proof of Lemma~\ref{SM-lemma:order-leader}]
By construction, $\bm Z_l(\bm\theta_0)\sim N_d(\bm0,\bm I_d)$, so $\|\bm Z_l(\bm\theta_0)\|^2\sim\chi_d^2$. Since $d$ is fixed, for sufficiently large $x$,
$$
\mathbb P\{\|\bm Z_l(\bm\theta_0)\|>x\} =\mathbb P\{\|\bm Z_l(\bm\theta_0)\|^2>x^2\} \lesssim x^{d-2}\exp(-x^2/2).
$$
Hence, by the union bound,
\begin{align*}
\mathbb P\left\{\max_{1\le l\le L}\|\bm Z_l(\bm\theta_0)\|>M\sqrt{\log L}\right\}
&\le L\,\mathbb P\{\|\bm Z_1(\bm\theta_0)\|>M\sqrt{\log L}\} \\
&\lesssim (\log L)^{(d-2)/2}L^{1-M^2/2}\to0
\end{align*}
for sufficiently large $M$, proving $\max_l\|\bm Z_l(\bm\theta_0)\| = O_p(\sqrt{\log L})$.

For the expectation, let $M_L:=\max_{1\le l\le L}\|\bm Z_l(\bm\theta_0)\|$ and $a_L=M\sqrt{\log L}$ for sufficiently large fixed $M$. Using $\mathbb E(M_L)=\int_0^\infty\mathbb P(M_L>x)\,dx$ and the same union bound,
$$
\mathbb E(M_L)\le a_L+CL\int_{a_L}^\infty x^{d-2}e^{-x^2/2}\,dx = O(\sqrt{\log L}),
$$
since the second term is $o(1)$ for sufficiently large $M$.
\end{proof}

\begin{proof}[Proof of Lemma~\ref{SM-lemma:order-reminder}]
Denote $\bm R_{l,n}:=\bm T_{l,n}-\bm Z_l(\bm\theta_0)$. Then
\begin{align*}
\bm R_{l,n}
&= \left\{\widehat{\bm\Sigma}_l^{-1/2}(\widehat{\bm\theta})-\bm G_{1,l}^{-1/2}(\bm\theta_0)\right\}
\bm K_l^\top\{\check{\bm f}(\bm\theta_0)-\bm f\} \\
&\quad+\widehat{\bm\Sigma}_l^{-1/2}(\widehat{\bm\theta})\bm K_l^\top
\{\widehat{\bm f}(\widehat{\bm\theta})-\check{\bm f}(\bm\theta_0)\}
+\widehat{\bm\Sigma}_l^{-1/2}(\widehat{\bm\theta})\bm C_l^\top(\widehat{\bm\beta}-\bm\beta_0) \\
&:= \bm R_{1,l,n}+\bm R_{2,l,n}+\bm R_{3,l,n}.
\end{align*}

For $\bm R_{1,l,n}$, Lemma~\ref{SM-lemma:order_of_G1andG2} gives
$$
\max_{1\le l\le L}\|\widehat{\bm\Sigma}_l(\widehat{\bm\theta})-\bm G_{1,l}(\bm\theta_0)\| \le \max_l\|\bm G_{1,l}(\widehat{\bm\theta})-\bm G_{1,l}(\bm\theta_0)\|+\max_l\|\bm G_{2,l}(\widehat{\bm\theta})\| = O_p(n^{-1/2}).
$$
By Assumption~\ref{SM-regcond:PMSE} and the local Lipschitz continuity of the inverse square-root map,
$$
\max_{1\le l\le L}\|\widehat{\bm\Sigma}_l^{-1/2}(\widehat{\bm\theta})-\bm G_{1,l}^{-1/2}(\bm\theta_0)\|=O_p(n^{-1/2}).
$$
Moreover, since $\bm Z_l(\bm\theta_0)=\bm G_{1,l}^{-1/2}(\bm\theta_0)\bm K_l^\top\{\check{\bm f}(\bm\theta_0)-\bm f\}\sim N_d(\bm0,\bm I_d)$,
$$
\max_{1\le l\le L}\|\bm K_l^\top\{\check{\bm f}(\bm\theta_0)-\bm f\}\| \le \max_l\|\bm G_{1,l}^{1/2}(\bm\theta_0)\|\max_l\|\bm Z_l(\bm\theta_0)\| = O_p(\sqrt{\log L}),
$$
where Lemma~\ref{SM-lemma:order-leader} and Assumption~\ref{SM-regcond:PMSE} are used in the last equality. Hence
$$
\max_{1\le l\le L}\|\bm R_{1,l,n}\|=O_p\left(\sqrt{\frac{\log L}{n}}\right).
$$

For $\bm R_{2,l,n}$, Assumption~\ref{SM-regcond:PMSE} and Lemma~\ref{SM-lemma:order_of_G1andG2} imply $\max_l\|\widehat{\bm\Sigma}_l^{-1/2}(\widehat{\bm\theta})\|=O_p(1)$. By the definitions of $\widehat{\bm f}(\widehat{\bm\theta})$ and $\check{\bm f}(\bm\theta_0)$,
\begin{align*}
\bm K_l^\top\{\widehat{\bm f}(\widehat{\bm\theta})-\check{\bm f}(\bm\theta_0)\}
&= \bm K_l^\top\left[\{(\widehat{\bm\Lambda}^\top\widehat{\bm V}^{-1}-\bm\Lambda_0^\top\bm V_0^{-1})\otimes\bm I_n\}
\{\bm y-(\bm I_p\otimes\bm X)\bm\beta_0\}\right] \\
&\quad+\bm K_l^\top\left[\{(\widehat{\bm\Lambda}^\top\widehat{\bm V}^{-1})\otimes\bm I_n\}
(\bm I_p\otimes\bm X)(\bm\beta_0-\widehat{\bm\beta})\right].
\end{align*}
Smoothness of $\bm\Lambda^\top\bm V^{-1}$ and $\|\widehat{\bm\theta}-\bm\theta_0\|=O_p(n^{-1/2})$ give
$\|\widehat{\bm\Lambda}^\top\widehat{\bm V}^{-1}-\bm\Lambda_0^\top\bm V_0^{-1}\|=O_p(n^{-1/2})$. Thus, by Assumption~\ref{SM-regcond:designed-mat},
$$
\max_l\left\|\bm K_l^\top\left[\{(\widehat{\bm\Lambda}^\top\widehat{\bm V}^{-1}-\bm\Lambda_0^\top\bm V_0^{-1})\otimes\bm I_n\} \{\bm y-(\bm I_p\otimes\bm X)\bm\beta_0\}\right]\right\| = O_p\left(\sqrt{\frac{\log L}{n}}\right),
$$
since $\max_{i\in\cup_l\mathcal S_l}\|\bm y_i-\bm B_0\bm x_i\|=O_p(\sqrt{\log L})$ and $|\cup_l\mathcal S_l|\le S_KL$. Furthermore, Assumptions~\ref{SM-regcond:designed-mat} and \ref{SM-regcond:X}, together with $\|\widehat{\bm\Lambda}^\top\widehat{\bm V}^{-1}\|=O_p(1)$ and $\|\widehat{\bm\beta}-\bm\beta_0\|=O_p(n^{-1/2})$, give
$$
\max_l\left\|\bm K_l^\top\left[\{(\widehat{\bm\Lambda}^\top\widehat{\bm V}^{-1})\otimes\bm I_n\} (\bm I_p\otimes\bm X)(\bm\beta_0-\widehat{\bm\beta})\right]\right\|=O_p(n^{-1/2}).
$$
Therefore,
$$
\max_{1\le l\le L}\|\bm R_{2,l,n}\|=O_p\left(\sqrt{\frac{\log L}{n}}\right).
$$

Finally,
$$
\max_{1\le l\le L}\|\bm R_{3,l,n}\| \le \max_l\|\widehat{\bm\Sigma}_l^{-1/2}(\widehat{\bm\theta})\| \max_l\|\bm C_l\|\|\widehat{\bm\beta}-\bm\beta_0\| = O_p(n^{-1/2}) = O_p\left(\sqrt{\frac{\log L}{n}}\right),
$$
by Assumptions~\ref{SM-regcond:designed-mat} and \ref{SM-regcond:PMSE}. Combining the three bounds yields
$$
\max_{1\le l\le L}\|\bm R_{l,n}\|=O_p\left(\sqrt{\frac{\log L}{n}}\right),
$$
as required.
\end{proof}

\newpage
\section{Additional methodological details}
\label{SM-sec:supp-algorithms}

\subsection{Choice of PMSE approximation}
\label{SM-sec:supp-pmse-choice}

This section provides additional discussion on the choice of PMSE approximation used to standardise prediction errors. Recall from Equation~\eqref{eq:g1g2} in Section~\ref{subsec:ppr} of the main text that, for the $l$th mixed target, the PMSE of the BLUP admits the decomposition $\bm\Sigma(\widetilde{\bm\mu}_l;\bm\theta)=\bm G_{1,l}(\bm\theta) + \bm G_{2,l}(\bm\theta)$, where $\bm G_{1,l}(\bm\theta)$ represents the leading uncertainty from predicting the latent component, and $\bm G_{2,l}(\bm\theta)$ represents the additional uncertainty induced by estimating the regression coefficients. Under Conditions \ref{SM-regcond:true-para}--\ref{SM-regcond:PMSE}, Assumption~\ref{SM-regcond:PMSE} ensures that $\bm G_{1,l}(\bm\theta_0)$ is non-degenerate, while $\bm G_{2,l}(\widehat{\bm\theta}) = O_p(n^{-1})$ (see Lemma~\ref{SM-lemma:order_of_G1andG2}). Thus, $\bm G_{2,l}(\widehat{\bm\theta})$ is asymptotically negligible relative to the leading prediction uncertainty.

In finite samples, however, the choice between $\bm G_{1,l}(\widehat{\bm\theta})$ and $\bm G_{1,l}(\widehat{\bm\theta})+\bm G_{2,l}(\widehat{\bm\theta})$ can affect the prediction regions in two ways. First, it changes the geometry of each marginal ellipsoid. Recall that for a generic PMSE matrix $\widehat{\bm\Sigma}_l$, we introduced the component region in Section~\ref{sec:bootstrap} having the form
$$
\mathcal S_l(c,\widehat{\bm\Sigma}_l) = \{\bm x\in\mathbb R^d:\dM(\widehat{\bm\mu}_l,\bm x;\widehat{\bm\Sigma}_l)\leq c\}, \quad |\mathcal S_l(c,\widehat{\bm\Sigma}_l)| = \frac{\pi^{d/2}}{\Gamma(d/2+1)}c^d|\widehat{\bm\Sigma}_l|^{1/2}.
$$
For a fixed critical value $c$ then, we see that replacing $\bm G_{1,l}(\widehat{\bm\theta})$ by $\bm G_{1,l}(\widehat{\bm\theta})+\bm G_{2,l}(\widehat{\bm\theta})$ cannot decrease the determinant and produces a region containing the original region, because $\bm G_{2,l}(\widehat{\bm\theta})$ is positive semidefinite.

On the other hand, recall that the same PMSE matrix is used to standardise when estimating the distribution of the maximum statistic. For example, the bootstrap SPR statistic has the form
$$
\tau^*(\widehat{\bm\Sigma}) = \max_{1\leq l\leq L} \dM(\widehat{\bm\mu}^{*}_l,\bm\mu^{*}_l;\widehat{\bm\Sigma}^{*}_l), \qquad
c^{\mathrm{SPR,BS}}_{1-\alpha}(\widehat{\bm\Sigma}) = q^*_{1-\alpha}\{\tau^*(\widehat{\bm\Sigma})\}.
$$
Therefore, changing the PMSE approximation affects not only the final ellipsoid volume, but also the estimated upper tail of the calibrated pivot:
a larger PMSE matrix can reduce the bootstrap or Monte Carlo statistics, which may in turn reduce the estimated critical value. Consequently, the overall effect on coverage is not determined solely by the marginal size of $\widehat{\bm\Sigma}_l$.

The above trade-off between the first and second points explains why the full approximation $\bm G_{1,l}(\bm\theta) + \bm G_{2,l}(\bm\theta)$ is not guaranteed to necessarily improve the finite-sample performance of the bootstrap or Monte Carlo SPR methods. Although it is a more complete approximation to the PMSE, it may distort the standardisation of the maximum statistic when $\bm G_{2,l}(\bm\theta)$ is estimated with non-negligible finite-sample error. Indeed, in our simulations using $\bm G_{1,l}(\widehat{\bm\theta})$ alone results in more stable behaviour of the bootstrap and Monte Carlo SPRs (see detailed simulation results in Section~\ref{SM-sec:supplement-simres}). This is also consistent with the asymptotic order of the decomposition: since $\bm G_{2,l}(\widehat{\bm\theta})$ is of smaller order, omitting it does not affect the leading approximation to the prediction error, and it may avoid additional finite-sample noise in the estimated studentisation.

On the other hand, the situation is different for the Bonferroni SPR method, where its critical value is analytic and does not require estimation:
$$
c^{\mathrm{SPR,BO}}_{1-\alpha} = \{F^{-1}_{\chi_d^2}(1-\alpha/L)\}^{1/2}.
$$
Therefore, changing the approximation to the PMSE affects only the geometry of the final regions, not the calibration of the critical value. In this case, we would anticipate that using only $\bm G_{1,l}(\widehat{\bm\theta})$ may lead to SPRs that are too small in finite samples, because the fixed Bonferroni critical value does not compensate for this. Indeed, it explains why the full approximation $\bm G_{1,l}(\widehat{\bm\theta})+\bm G_{2,l}(\widehat{\bm\theta})$ is preferable for the Bonferroni method in our numerical studies (see Section~\ref{SM-sec:supplement-simres}).

\subsection{Connection between Bonferroni critical value and Gumbel calibration}
\label{SM-sec:supp-gumbel-bonferroni}

This section explains the connection between the Bonferroni critical value and the Gumbel approximation for the maximum statistic. The argument is an idealised benchmark: in the FA model setting, the Mahalanobis statistics are generally dependent and involve estimated variance components. Nevertheless, it helps explain why the Bonferroni method can be interpreted as a simple analytic approximation to the upper tail of the maximum statistic.

For the $l$th general mixed parameter target, write the squared Mahalanobis statistic as $Q_l=\dM^2(\widehat{\bm\mu}_l,\bm\mu_l;\widehat{\bm\Sigma}(\widehat{\bm\mu}_l))$. Under the pointwise normal approximation stated in Proposition~\ref{SM-prop:normal-approx}, $Q_l$ is approximately distributed as $\chi_d^2$. If $Q_1,\ldots,Q_L$ are further treated as independent variables, then the squared maximum statistic $M_L=\max_{1\leq l\leq L}Q_l$ satisfies
$$
\mathbb P(M_L\leq t)=\{F_{\chi_d^2}(t)\}^L, \quad c^{2}_{1-\alpha} = F^{-1}_{\chi_d^2}\{(1-\alpha)^{1/L}\}.
$$
Since $(1-\alpha)^{1/L} = 1-\{-\log(1-\alpha)\}/L+O(L^{-2})$, then the independent benchmark is approximately the marginal chi-square quantile with tail probability $-\log(1-\alpha)/L$. Because $-\log(1-\alpha)\approx\alpha$ for small $\alpha$, then it follows that the Bonferroni squared critical value $c^{2,\mathrm{BO}}_{1-\alpha}=F^{-1}_{\chi_d^2}(1-\alpha/L)$ is close to the exact independent maximum-based critical value e.g., $-\log(0.95)\approx0.0513$ when $\alpha=0.05$.

The same relationship can be derived from extreme-value theory. Let $X_1,\ldots,X_L$ be independent $\chi_d^2$ random variables, and define $M_L=\max_{1\leq l\leq L}X_l$. The upper tail satisfies
$$ 
1-F_{\chi_d^2}(t) \sim \frac{1}{2^{d/2-1}\Gamma(d/2)}t^{d/2-1}\exp\left(-\frac{t}{2}\right), \quad t\to\infty.
$$
Setting $b_L=2\log L+(d-2)\log\log L-2\log\Gamma(d/2)$ then, this tail approximation gives
$L\{1-F_{\chi_d^2}(b_L+2x)\}\to \exp(-x)$, and $\mathbb P(M_L\leq b_L+2x)\to \exp\{-\exp(-x)\}$. Thus $(M_L-b_L)/2$ converges to a standard Gumbel distribution. Finally, solving $\exp\{-\exp(-x_\alpha)\}=1-\alpha$ gives $x_\alpha=-\log\{-\log(1-\alpha)\}$, and so the Gumbel squared critical value is given by
$$
c^{2,\mathrm{Gum}}_{1-\alpha} = b_L-2\log\{-\log(1-\alpha)\}.
$$
On the other hand, the Bonferroni quantile is determined by $1-F_{\chi_d^2}(c^{2,\mathrm{BO}}_{1-\alpha})=\alpha/L$. Applying the same chi-square tail approximation yields
$$
c^{2,\mathrm{BO}}_{1-\alpha} = 2\log L+(d-2)\log\log L-2\log\Gamma(d/2)-2\log\alpha+o(1).
$$
Therefore, the Bonferroni and Gumbel critical values have the same leading extreme-value form, with the only difference being the replacement of $-\log(1-\alpha)$ by $\alpha$.

This derivation also clarifies why the Bonferroni method often behaves like an extreme-value approximation to the maximum statistic and hence its limitations. The argument relies on accurate marginal chi-square calibration and an idealised independence, or sufficiently weak-dependence structure. In FA models, the Mahalanobis statistics are functions of the same estimated loading matrix, uniquenesses, regression coefficients, and latent factors, and their finite-sample behaviour also depends on the PMSE approximation. Thus, although the Bonferroni method provides a simple analytic approximation, bootstrap calibration is generally expected to be more reliable in finite samples.

\subsection{Detailed algorithms}
\label{SM-sec:detailed-algorithm}

In this section, we provide formal algorithms used to construct SPRs by using the bootstrap procedure (Algorithm~\ref{SM-alg:bootstrap_spr}), the Monte Carlo procedure (Algorithm~\ref{SM-alg:mc-spr}), and the Bonferroni procedure (Algorithm~\ref{SM-alg:bo-spr}).

\begin{algorithm}[!htbp]
\caption{Bootstrap SPRs for General Mixed Parameters}
\label{SM-alg:bootstrap_spr}

\begin{algorithmic} [1]
\Require Observation $\boldsymbol{y}$, covariate matrix $\bm{X}$, designed matrices $\boldsymbol{C}_l$ and $\boldsymbol{K}_l$ for all $l=1,2,\dots, L$, confidence level $1-\alpha$, and the number of bootstrap $B$.

\State Fit the FA model to obtain parameter estimates $(\widehat{\boldsymbol{\theta}}^\top, \widehat{\boldsymbol{\beta}}^\top)^\top$, predicted mixed parameters $\widehat{\boldsymbol{\mu}}_l$, and the PMSE $\widehat{\bm{\Sigma}}(\widehat{\bm{\mu}}_l)$ for all $l=1,2,\dots, L$.

\For{$b = 1$ to $B$}
    \State Generate bootstrap factor $\boldsymbol{f}^{*(b)}$ from $\mathcal{N}_{nm}(\bm{0}_{nm}, \bm{I}_{nm})$ and bootstrap error $\boldsymbol{\epsilon}^{*(b)}$ from $\mathcal{N}_{np}(\bm{0}_{np}, \widehat{\boldsymbol{\Psi}} \otimes \bm{I}_n)$. Then generate the bootstrap response $\boldsymbol{y}^{*(b)} = (\bm{I}_p \otimes \bm{X})\widehat{\boldsymbol{\beta}} + (\widehat{\boldsymbol{\Lambda}} \otimes \bm{I}_n)\boldsymbol{f}^{*(b)} + \boldsymbol{\epsilon}^{*(b)}$ and the bootstrap underlying true general mixed parameters $\boldsymbol{\mu}^{*(b)}_l = \boldsymbol{C}_l^\top \widehat{\boldsymbol{\beta}} + \boldsymbol{K}_l^\top \boldsymbol{f}^{*(b)}$.
    
    \State Refit the FA model to $\boldsymbol y^{*(b)}$ to obtain $\widehat{\bm\theta}^{*(b)}$, $\widehat{\bm\beta}^{*(b)}$, and the refitted factor predictors.
    \State Align the refitted factor predictors and loading columns with the generated factors as in Section~\ref{SM-subsec:align-bootstrap}. Then compute $\widehat{\bm\mu}_l^{*(b)}$ and $\widehat{\bm\Sigma}^{*(b)}(\widehat{\bm\mu}_l^{*(b)})$ from the aligned quantities for every $l$.
    
    \State Compute the maximum of Mahalanobis distance $\tau^{*(b)} = \max_l \dM(\widehat{\bm{\mu}}^{*(b)}_l, \bm{\mu}^{*(b)}_l; \widehat{\bm{\Sigma}}^{*(b)}(\widehat{\bm{\mu}}_l^{*(b)}))$.
\EndFor

\State Compute the critical value $\widehat{c}^{\mathrm{SPR,BS}}_{1-\alpha}$ by using the $(1-\alpha)$-th quantile of $\{\tau^{*(1)}, \tau^{*(2)}, \dots, \tau^{*(B)}\}$.
\State Construct the $\widehat{\mathcal{S}}_{1-\alpha}^{\mathrm{BS}}$ using $\widehat{c}^{\mathrm{SPR,BS}}_{1-\alpha}$ according to Equation~\eqref{eq:spr-bs}.
\State \textbf{Return} Bootstrap SPR $\widehat{\mathcal{S}}_{1-\alpha}^{\mathrm{BS}}$.

\end{algorithmic}
\end{algorithm}

\begin{algorithm}[!htbp]
\caption{Monte Carlo SPRs for General Mixed Parameters}
\label{SM-alg:mc-spr}

\begin{algorithmic}[1]
\Require Observation $\bm y$, covariate matrix $\bm X$, target-design matrices $\bm C_l$ and $\bm K_l$ for $l=1,\ldots,L$, confidence level $1-\alpha$, and number of Monte Carlo samples $S$.

\State Fit the FA model to obtain $(\widehat{\bm\theta}^\top,\widehat{\bm\beta}^\top)^\top$, predicted mixed parameters $\widehat{\bm\mu}_l$, and PMSE matrices $\widehat{\bm\Sigma}(\widehat{\bm\mu}_l)$ for all $l=1,\ldots,L$.
\State Compute the plug-in precision matrix $\widehat{\bm\Omega}_{\xi} = \widehat{\bm A}^\top\widehat{\bm W}^{-1}\widehat{\bm A}+\bm P$.

\For{$s=1$ to $S$}
    \State Draw $\bm u^{(s)}\sim\mathcal N_{qp+nm}(\bm0_{qp+nm}, \widehat{\bm\Omega}_{\xi}^{-1})$.
    \State For each $l=1,\ldots,L$, compute $\tau_l^{(s)} = \{(\bm u^{(s)})^\top \bm H_l^\top \widehat{\bm\Sigma}(\widehat{\bm\mu}_l)^{-1} \bm H_l \bm u^{(s)} \}^{1/2}$ defined in Section~\ref{subsec:MC-method}.
    \State Compute $\tau^{(s)}=\max_{1\leq l\leq L}\tau_l^{(s)}$.
\EndFor

\State Compute $\widehat{c}^{\mathrm{SPR,MC}}_{1-\alpha}$ as the empirical $(1-\alpha)$-quantile of $\{\tau^{(1)},\ldots,\tau^{(S)}\}$.
\State Construct $\widehat{\mathcal{S}}_{1-\alpha}^{\mathrm{MC}}$ using $\widehat{c}^{\mathrm{SPR,MC}}_{1-\alpha}$ by the Monte-Carlo analogue of Equation~\eqref{eq:spr-bs}.
\State \textbf{Return} Monte-Carlo SPR $\widehat{\mathcal{S}}_{1-\alpha}^{\mathrm{MC}}$.

\end{algorithmic} 
\end{algorithm}

\begin{algorithm}[!htbp]
\caption{Bonferroni SPRs for General Mixed Parameters}
\label{SM-alg:bo-spr}

\begin{algorithmic}[1]
\Require Observation $\bm y$, covariate matrix $\bm X$, target-design matrices $\bm C_l$ and $\bm K_l$ for $l=1,\ldots,L$, confidence level $1-\alpha$.

\State Fit the FA model to obtain $(\widehat{\bm\theta}^\top,\widehat{\bm\beta}^\top)^\top$, predicted mixed parameters $\widehat{\bm\mu}_l$, and PMSE matrices $\widehat{\bm\Sigma}(\widehat{\bm\mu}_l)$ for all $l=1,\ldots,L$.

\State Compute $c^{\mathrm{SPR,BO}}_{1-\alpha} = \{F^{-1}_{\chi^2_d}(1-L^{-1}\alpha)\}^{1/2}$.
\State Construct $\mathcal S_{1-\alpha}^{\mathrm{BO}}$ using $c^{\mathrm{SPR,BO}}_{1-\alpha}$ by the Bonferroni analogue of Equation~\eqref{eq:spr-bs}.
\State \textbf{Return} Bonferroni SPR $\mathcal S_{1-\alpha}^{\mathrm{BO}}$.

\end{algorithmic}
\end{algorithm}

\FloatBarrier
\subsection{Additional processing for bootstrap SPRs}
\label{SM-subsec:align-bootstrap}

Although the corner constraint removes the main rotational indeterminacy in the loading matrix, finite sample instability may still arise when comparing estimates obtained from FA models fitted to different bootstrap samples. In particular, the predicted factor scores may exhibit sign or order switching across bootstrap replications. If left uncorrected, then it may result in the bootstrap predictors $\widehat{\bm\mu}^{*(b)}_l$ having a different factor orientation relative to the target $\bm\mu^{*(b)}_l$, meaning the resulting bootstrap prediction error $\widehat{\bm\mu}^{*(b)}_l - \bm\mu^{*(b)}_l$ would no longer correspond to the intended latent target.

To rectify this issue, we propose to align the refitted bootstrap predicted scores and loadings with the generated bootstrap scores before computing the Mahalanobis statistic. In detail, for the $b$th bootstrap sample, let $\bm f^{*(b)}_k=(f^{*(b)}_{1k},\ldots,f^{*(b)}_{nk})^\top$ denote the generated bootstrap scores for factor $k$, and let $\widehat{\bm f}^{*(b)}_j$ denote the corresponding refitted bootstrap predictor for factor $j$. Then we compute the absolute Pearson correlations $|\operatorname{cor}(\widehat{\bm f}^{*(b)}_j,\bm f^{*(b)}_k)|$ for all $j,k=1,\ldots,m$, and choose the permutation $\widehat\pi^{*(b)}$ that gives the largest total matched correlation, i.e.,
$$
\widehat\pi^{*(b)} = \arg\max_{\pi\in\mathfrak S_m} \sum_{k=1}^m
|\operatorname{cor}(\widehat{\bm f}^{*(b)}_{\pi(k)},\bm f^{*(b)}_k)|,
$$
where $\mathfrak S_m$ denotes the set of all permutations of $\{1,\ldots,m\}$. The refitted bootstrap factor predictors and the columns of the refitted loading matrix are then reordered according to $\widehat\pi^{*(b)}$. 

After this reordering, we attempt to correct for any remaining sign indeterminacy by defining $s_k^{*(b)}=\operatorname{sign}\{\operatorname{cor}(\widehat{\bm f}^{*(b)}_{\widehat\pi^{*(b)}(k)},\bm f^{*(b)}_k)\}$ and replacing the aligned $k$th refitted factor predictor and loading column by $s_k^{*(b)}\widehat{\bm f}^{*(b)}_{\widehat\pi^{*(b)}(k)}$ and $s_k^{*(b)}\widehat{\bm\lambda}^{*(b)}_{\widehat\pi^{*(b)}(k)}$, respectively. With this processing, it is hoped that $\widehat{\bm\mu}^{*(b)}_l$ and $\bm\mu^{*(b)}_l$ can be compared under the same factor ordering and sign convention. 

We reiterate that the above processing step should be understood as a within-bootstrap correction: it ensures each refitted bootstrap predicted score is expressed in the same factor orientation as the generated bootstrap target. However, it does not address possible global misspecification or a poor initial FA fit. Put another way, as with standard parametric bootstrap procedures, such alignment is conditional on the fitted model. We also acknowledge that future work may look into (more sophisticated) variations of the above processing approach e.g., using Spearman rank correlations, directly employ a direct matrix rotation such as Procrustes or varimax rotation, and so on. Such variations may be particularly useful if the number of latent variables $m$ becomes non-negligible. 

\newpage 
\FloatBarrier

\begin{sidewaystable}[p]
\section{Additional simulation results}
\label{SM-sec:supplement-simres}
\centering
\caption{Simulation results using the leading PMSE component $\bm G_1$ for $p=10$.}
\label{SM-tab:simulation-G1-p10}
\resizebox{\linewidth}{!}{%
\begin{tabular}{c|rrrrr|rrrrr|rrrrr}
\hline
& \multicolumn{5}{c|}{ECP}
& \multicolumn{5}{c|}{AV}
& \multicolumn{5}{c}{AVV} \\
Method
& 50 & 100 & 200 & 500 & 1000
& 50 & 100 & 200 & 500 & 1000
& 50 & 100 & 200 & 500 & 1000 \\ \hline

\multicolumn{16}{c}{$\mathrm{VE}=0.3$} \\ \hline
BS
& 0.884 & 0.974 & 0.969 & 0.938 & 0.961
& 650.250 & 389.117 & 38.178 & 20.695 & 21.448
& 205695.69979 & 250983.69592 & 14985.31908 & 1.44825 & 0.75183 \\
MC
& 0.264 & 0.582 & 0.803 & 0.890 & 0.933
& 15.517 & 14.770 & 16.767 & 18.972 & 20.524
& 27.45767 & 2.69357 & 1.73004 & 1.01984 & 0.62417 \\
BO
& 0.133 & 0.482 & 0.779 & 0.883 & 0.928
& 8.054 & 13.354 & 16.228 & 18.778 & 20.466
& 16.83125 & 6.30217 & 1.85176 & 0.95540 & 0.54289 \\
PPR
& 0.000 & 0.000 & 0.000 & 0.000 & 0.000
& 3.493 & 5.263 & 5.861 & 6.108 & 6.191
& 3.16555 & 0.97896 & 0.24158 & 0.10107 & 0.04968 \\ \hline

\multicolumn{16}{c}{$\mathrm{VE}=0.5$} \\ \hline
BS
& 0.977 & 0.979 & 0.952 & 0.936 & 0.956
& 116.764 & 14.923 & 11.531 & 11.214 & 11.556
& 18750.11659 & 152.83319 & 0.84869 & 0.35754 & 0.20005 \\
MC
& 0.651 & 0.776 & 0.870 & 0.907 & 0.937
& 10.149 & 9.457 & 9.704 & 10.459 & 11.153
& 2.69762 & 0.74466 & 0.51755 & 0.30184 & 0.17816 \\
BO
& 0.251 & 0.557 & 0.781 & 0.883 & 0.925
& 6.370 & 7.874 & 8.955 & 10.162 & 11.020
& 1.99535 & 0.84152 & 0.51545 & 0.28435 & 0.15724 \\
PPR
& 0.001 & 0.000 & 0.000 & 0.000 & 0.000
& 2.763 & 3.103 & 3.235 & 3.305 & 3.334
& 0.37528 & 0.13072 & 0.06725 & 0.03008 & 0.01439 \\ \hline

\multicolumn{16}{c}{$\mathrm{VE}=0.7$} \\ \hline
BS
& 0.990 & 0.965 & 0.952 & 0.949 & 0.944
& 15.046 & 8.076 & 6.358 & 5.718 & 5.715
& 46.932112 & 0.886848 & 0.183791 & 0.080755 & 0.045231 \\
MC
& 0.842 & 0.833 & 0.856 & 0.914 & 0.914
& 7.798 & 5.969 & 5.331 & 5.298 & 5.498
& 0.573194 & 0.198939 & 0.122631 & 0.070085 & 0.040418 \\
BO
& 0.147 & 0.379 & 0.640 & 0.843 & 0.882
& 3.324 & 3.887 & 4.356 & 4.912 & 5.315
& 0.346836 & 0.197941 & 0.123094 & 0.066633 & 0.036244 \\
PPR
& 0.000 & 0.000 & 0.000 & 0.000 & 0.000
& 1.442 & 1.532 & 1.573 & 1.598 & 1.608
& 0.065231 & 0.030748 & 0.016059 & 0.007049 & 0.003316 \\ \hline

\multicolumn{16}{c}{$\mathrm{VE}=0.9$} \\ \hline
BS
& 0.980 & 0.957 & 0.959 & 0.956 & 0.948
& 9.495 & 5.151 & 3.249 & 2.219 & 1.885
& 1.0562319 & 0.1534935 & 0.0340865 & 0.0079511 & 0.0038736 \\
MC
& 0.905 & 0.875 & 0.884 & 0.894 & 0.897
& 6.497 & 3.830 & 2.576 & 1.943 & 1.744
& 0.4758350 & 0.0764638 & 0.0194124 & 0.0061764 & 0.0032258 \\
BO
& 0.005 & 0.041 & 0.146 & 0.488 & 0.704
& 0.956 & 1.097 & 1.220 & 1.372 & 1.482
& 0.0270569 & 0.0160059 & 0.0096346 & 0.0051604 & 0.0027961 \\
PPR
& 0.000 & 0.000 & 0.000 & 0.000 & 0.000
& 0.415 & 0.433 & 0.441 & 0.446 & 0.448
& 0.0050887 & 0.0024863 & 0.0012569 & 0.0005459 & 0.0002558 \\ \hline

\end{tabular}%
}
\par\smallskip
\footnotesize
BS, MC, BO, and PPR denote the bootstrap SPR, Monte Carlo SPR,
Bonferroni SPR, and pointwise prediction region, respectively.
ECP, AV, and AVV denote the empirical simultaneous coverage probability,
average volume, and average variance of volume, respectively.
All procedures use the leading PMSE component $\bm G_1$ only.
\end{sidewaystable}

\begin{sidewaystable}[p]
\centering
\caption{Simulation results using the leading PMSE component $\bm G_1$ for $p=20$.}
\label{SM-tab:simulation-G1-p20}
\resizebox{\linewidth}{!}{%
\begin{tabular}{c|rrrrr|rrrrr|rrrrr}
\hline
& \multicolumn{5}{c|}{ECP}
& \multicolumn{5}{c|}{AV}
& \multicolumn{5}{c}{AVV} \\
Method
& 50 & 100 & 200 & 500 & 1000
& 50 & 100 & 200 & 500 & 1000
& 50 & 100 & 200 & 500 & 1000 \\ \hline

\multicolumn{16}{c}{$\mathrm{VE}=0.3$} \\ \hline
BS
& 0.977 & 0.970 & 0.949 & 0.960 & 0.957
& 31.378 & 14.724 & 12.528 & 12.066 & 12.404
& 3759.30947 & 5.13463 & 1.16706 & 0.36886 & 0.20563 \\
MC
& 0.603 & 0.742 & 0.846 & 0.916 & 0.938
& 10.071 & 9.750 & 10.196 & 11.113 & 11.893
& 1.41935 & 0.72641 & 0.52135 & 0.28694 & 0.17770 \\
BO
& 0.187 & 0.549 & 0.762 & 0.898 & 0.920
& 6.757 & 8.368 & 9.525 & 10.853 & 11.777
& 1.02194 & 0.77384 & 0.51793 & 0.26882 & 0.15542 \\
PPR
& 0.001 & 0.000 & 0.000 & 0.000 & 0.000
& 2.930 & 3.298 & 3.440 & 3.530 & 3.563
& 0.19220 & 0.12021 & 0.06757 & 0.02844 & 0.01422 \\ \hline

\multicolumn{16}{c}{$\mathrm{VE}=0.5$} \\ \hline
BS
& 0.982 & 0.958 & 0.959 & 0.964 & 0.950
& 12.413 & 8.220 & 6.629 & 6.058 & 6.083
& 2.948158 & 0.619814 & 0.205240 & 0.079391 & 0.047910 \\
MC
& 0.819 & 0.825 & 0.867 & 0.912 & 0.928
& 7.765 & 5.955 & 5.491 & 5.592 & 5.849
& 0.269293 & 0.186852 & 0.124283 & 0.067648 & 0.041571 \\
BO
& 0.133 & 0.429 & 0.673 & 0.862 & 0.908
& 3.546 & 4.166 & 4.659 & 5.258 & 5.687
& 0.285965 & 0.205402 & 0.127274 & 0.063138 & 0.035825 \\
PPR
& 0.000 & 0.000 & 0.000 & 0.000 & 0.000
& 1.538 & 1.642 & 1.683 & 1.710 & 1.720
& 0.053783 & 0.031907 & 0.016604 & 0.006680 & 0.003278 \\ \hline

\multicolumn{16}{c}{$\mathrm{VE}=0.7$} \\ \hline
BS
& 0.979 & 0.963 & 0.957 & 0.959 & 0.951
& 9.453 & 5.495 & 3.865 & 3.088 & 2.920
& 0.640810 & 0.138594 & 0.043373 & 0.016484 & 0.010020 \\
MC
& 0.903 & 0.874 & 0.874 & 0.892 & 0.924
& 6.734 & 4.124 & 3.151 & 2.795 & 2.774
& 0.186856 & 0.050718 & 0.026512 & 0.014434 & 0.008563 \\
BO
& 0.028 & 0.149 & 0.389 & 0.722 & 0.856
& 1.659 & 1.912 & 2.123 & 2.386 & 2.577
& 0.065595 & 0.043998 & 0.026733 & 0.012964 & 0.007306 \\
PPR
& 0.000 & 0.000 & 0.000 & 0.000 & 0.000
& 0.719 & 0.754 & 0.767 & 0.776 & 0.779
& 0.012337 & 0.006834 & 0.003488 & 0.001372 & 0.000668 \\ \hline

\multicolumn{16}{c}{$\mathrm{VE}=0.9$} \\ \hline
BS
& 0.970 & 0.970 & 0.960 & 0.948 & 0.951
& 8.112 & 4.076 & 2.358 & 1.416 & 1.071
& 0.35603458 & 0.05449446 & 0.01132539 & 0.00229341 & 0.00095671 \\
MC
& 0.925 & 0.899 & 0.901 & 0.886 & 0.898
& 6.202 & 3.158 & 1.873 & 1.212 & 0.963
& 0.17342479 & 0.02568395 & 0.00550595 & 0.00169303 & 0.00070949 \\
BO
& 0.000 & 0.002 & 0.027 & 0.193 & 0.452
& 0.452 & 0.515 & 0.570 & 0.639 & 0.690
& 0.00496546 & 0.00320301 & 0.00194777 & 0.00092879 & 0.00052012 \\
PPR
& 0.000 & 0.000 & 0.000 & 0.000 & 0.000
& 0.196 & 0.203 & 0.206 & 0.208 & 0.209
& 0.00093388 & 0.00049755 & 0.00025410 & 0.00009826 & 0.00004759 \\ \hline

\end{tabular}%
}
\par\smallskip
\footnotesize
BS, MC, BO, and PPR denote the bootstrap SPR, Monte Carlo SPR,
Bonferroni SPR, and pointwise prediction region, respectively.
ECP, AV, and AVV denote the empirical simultaneous coverage probability,
average volume, and average variance of volume, respectively.
All procedures use the leading PMSE component $\bm G_1$ only.
\end{sidewaystable}

\begin{sidewaystable}[p]
\centering
\caption{Simulation results using the full PMSE $\bm G_1+\bm G_2$ for $p=10$.}
\label{SM-tab:simulation-G1G2-p10}
\resizebox{\linewidth}{!}{%
\begin{tabular}{c|rrrrr|rrrrr|rrrrr}
\hline
& \multicolumn{5}{c|}{ECP}
& \multicolumn{5}{c|}{AV}
& \multicolumn{5}{c}{AVV} \\
Method
& 50 & 100 & 200 & 500 & 1000
& 50 & 100 & 200 & 500 & 1000
& 50 & 100 & 200 & 500 & 1000 \\ \hline

\multicolumn{16}{c}{$\mathrm{VE}=0.3$} \\ \hline
BS
& 0.883 & 0.966 & 0.966 & 0.941 & 0.931
& 109.785 & 104.385 & 27.779 & 20.666 & 21.368
& 3348.77347 & 11240.89066 & 1788.04702 & 1.45183 & 0.68788 \\
MC
& 0.237 & 0.568 & 0.809 & 0.894 & 0.906
& 10.541 & 14.399 & 16.723 & 18.962 & 20.458
& 11.13668 & 5.29693 & 1.82283 & 1.02093 & 0.57192 \\
BO
& 0.239 & 0.575 & 0.813 & 0.896 & 0.914
& 10.676 & 14.479 & 16.801 & 19.028 & 20.539
& 10.58524 & 5.18152 & 1.76888 & 0.94450 & 0.50600 \\
PPR
& 0.001 & 0.000 & 0.000 & 0.000 & 0.000
& 4.630 & 5.707 & 6.068 & 6.189 & 6.213
& 1.99082 & 0.80488 & 0.23077 & 0.09992 & 0.04630 \\ \hline

\multicolumn{16}{c}{$\mathrm{VE}=0.5$} \\ \hline
BS
& 0.963 & 0.972 & 0.958 & 0.941 & 0.956
& 28.271 & 13.131 & 11.302 & 11.167 & 11.539
& 323.41755 & 11.32709 & 0.84349 & 0.35696 & 0.19953 \\
MC
& 0.546 & 0.753 & 0.872 & 0.904 & 0.933
& 8.660 & 9.102 & 9.610 & 10.434 & 11.146
& 1.65844 & 0.81361 & 0.52571 & 0.30161 & 0.17881 \\
BO
& 0.557 & 0.765 & 0.871 & 0.901 & 0.940
& 8.751 & 9.157 & 9.654 & 10.472 & 11.187
& 1.60207 & 0.79547 & 0.50273 & 0.28141 & 0.15645 \\
PPR
& 0.014 & 0.003 & 0.000 & 0.000 & 0.000
& 3.795 & 3.609 & 3.487 & 3.406 & 3.384
& 0.30131 & 0.12357 & 0.06559 & 0.02977 & 0.01432 \\ \hline

\multicolumn{16}{c}{$\mathrm{VE}=0.7$} \\ \hline
BS
& 0.957 & 0.960 & 0.959 & 0.949 & 0.935
& 9.736 & 6.984 & 5.999 & 5.627 & 5.672
& 2.281224 & 0.421333 & 0.179305 & 0.082090 & 0.042389 \\
MC
& 0.647 & 0.803 & 0.853 & 0.913 & 0.916
& 5.749 & 5.240 & 5.097 & 5.239 & 5.466
& 0.347192 & 0.200610 & 0.126319 & 0.071146 & 0.040026 \\
BO
& 0.674 & 0.818 & 0.865 & 0.910 & 0.914
& 5.865 & 5.295 & 5.130 & 5.258 & 5.487
& 0.326156 & 0.193485 & 0.121358 & 0.066191 & 0.033736 \\
PPR
& 0.026 & 0.004 & 0.000 & 0.000 & 0.000
& 2.543 & 2.087 & 1.853 & 1.710 & 1.660
& 0.061342 & 0.030055 & 0.015832 & 0.007003 & 0.003087 \\ \hline

\multicolumn{16}{c}{$\mathrm{VE}=0.9$} \\ \hline
BS
& 0.949 & 0.952 & 0.954 & 0.948 & 0.929
& 5.996 & 3.770 & 2.634 & 1.979 & 1.800
& 0.1233109 & 0.0440783 & 0.0173861 & 0.0070299 & 0.0036884 \\
MC
& 0.649 & 0.719 & 0.795 & 0.882 & 0.881
& 3.349 & 2.459 & 1.987 & 1.722 & 1.672
& 0.0443620 & 0.0213269 & 0.0115185 & 0.0060702 & 0.0033727 \\
BO
& 0.697 & 0.766 & 0.816 & 0.885 & 0.882
& 3.578 & 2.559 & 2.029 & 1.737 & 1.680
& 0.0296370 & 0.0172645 & 0.0100523 & 0.0052403 & 0.0028131 \\
PPR
& 0.054 & 0.011 & 0.002 & 0.000 & 0.000
& 1.552 & 1.008 & 0.733 & 0.565 & 0.508
& 0.0055740 & 0.0026818 & 0.0013114 & 0.0005544 & 0.0002574 \\ \hline

\end{tabular}%
}
\par\smallskip
\footnotesize
BS, MC, BO, and PPR denote the bootstrap SPR, Monte Carlo SPR,
Bonferroni SPR, and pointwise prediction region, respectively.
ECP, AV, and AVV denote the empirical simultaneous coverage probability,
average volume, and average variance of volume, respectively.
All procedures use the full PMSE $\bm G_1+\bm G_2$.
\end{sidewaystable}

\begin{sidewaystable}[p]
\centering
\caption{Simulation results using the full PMSE $\bm G_1+\bm G_2$ for $p=20$.}
\label{SM-tab:simulation-G1G2-p20}
\resizebox{\linewidth}{!}{%
\begin{tabular}{c|rrrrr|rrrrr|rrrrr}
\hline
& \multicolumn{5}{c|}{ECP}
& \multicolumn{5}{c|}{AV}
& \multicolumn{5}{c}{AVV} \\
Method
& 50 & 100 & 200 & 500 & 1000
& 50 & 100 & 200 & 500 & 1000
& 50 & 100 & 200 & 500 & 1000 \\ \hline

\multicolumn{16}{c}{$\mathrm{VE}=0.3$} \\ \hline
BS
& 0.961 & 0.961 & 0.942 & 0.956 & 0.959
& 20.200 & 14.188 & 12.379 & 12.035 & 12.391
& 85.06799 & 5.16225 & 1.16897 & 0.36872 & 0.20518 \\
MC
& 0.482 & 0.714 & 0.848 & 0.916 & 0.933
& 8.927 & 9.516 & 10.110 & 11.099 & 11.887
& 0.91745 & 0.77022 & 0.51938 & 0.28781 & 0.17845 \\
BO
& 0.498 & 0.726 & 0.847 & 0.916 & 0.936
& 9.003 & 9.569 & 10.172 & 11.138 & 11.930
& 0.86281 & 0.73139 & 0.50306 & 0.26574 & 0.15449 \\
PPR
& 0.004 & 0.000 & 0.000 & 0.000 & 0.000
& 3.904 & 3.771 & 3.674 & 3.623 & 3.609
& 0.16227 & 0.11361 & 0.06563 & 0.02811 & 0.01414 \\ \hline

\multicolumn{16}{c}{$\mathrm{VE}=0.5$} \\ \hline
BS
& 0.958 & 0.958 & 0.948 & 0.955 & 0.943
& 10.388 & 7.556 & 6.412 & 6.003 & 6.059
& 3.343335 & 0.620254 & 0.205439 & 0.081131 & 0.044050 \\
MC
& 0.632 & 0.773 & 0.854 & 0.915 & 0.925
& 5.873 & 5.443 & 5.344 & 5.558 & 5.826
& 0.286712 & 0.210983 & 0.129585 & 0.068704 & 0.041204 \\
BO
& 0.650 & 0.778 & 0.854 & 0.918 & 0.927
& 5.978 & 5.493 & 5.380 & 5.577 & 5.848
& 0.260707 & 0.195418 & 0.123808 & 0.062451 & 0.034219 \\
PPR
& 0.013 & 0.000 & 0.000 & 0.000 & 0.000
& 2.592 & 2.165 & 1.943 & 1.814 & 1.769
& 0.049033 & 0.030356 & 0.016152 & 0.006607 & 0.003131 \\ \hline

\multicolumn{16}{c}{$\mathrm{VE}=0.7$} \\ \hline
BS
& 0.953 & 0.958 & 0.961 & 0.954 & 0.950
& 7.079 & 4.648 & 3.553 & 2.991 & 2.890
& 0.479283 & 0.112540 & 0.041948 & 0.016721 & 0.009182 \\
MC
& 0.665 & 0.747 & 0.831 & 0.897 & 0.926
& 4.035 & 3.235 & 2.851 & 2.710 & 2.746
& 0.080540 & 0.049450 & 0.028269 & 0.014751 & 0.008764 \\
BO
& 0.698 & 0.775 & 0.839 & 0.899 & 0.929
& 4.193 & 3.303 & 2.881 & 2.722 & 2.757
& 0.061565 & 0.042378 & 0.026112 & 0.012840 & 0.007265 \\
PPR
& 0.032 & 0.007 & 0.000 & 0.000 & 0.000
& 1.818 & 1.302 & 1.041 & 0.885 & 0.834
& 0.011579 & 0.006583 & 0.003407 & 0.001358 & 0.000665 \\ \hline

\multicolumn{16}{c}{$\mathrm{VE}=0.9$} \\ \hline
BS
& 0.952 & 0.953 & 0.957 & 0.958 & 0.946
& 5.827 & 3.219 & 1.984 & 1.223 & 0.987
& 0.10042451 & 0.02162761 & 0.00576613 & 0.00157788 & 0.00079701 \\
MC
& 0.583 & 0.641 & 0.720 & 0.792 & 0.861
& 2.723 & 1.787 & 1.286 & 0.968 & 0.868
& 0.01653121 & 0.00659234 & 0.00283876 & 0.00127866 & 0.00072422 \\
BO
& 0.678 & 0.724 & 0.770 & 0.808 & 0.872
& 3.041 & 1.941 & 1.350 & 0.986 & 0.876
& 0.00489857 & 0.00318581 & 0.00179054 & 0.00092614 & 0.00051870 \\
PPR
& 0.076 & 0.030 & 0.003 & 0.000 & 0.000
& 1.319 & 0.765 & 0.488 & 0.321 & 0.265
& 0.00092130 & 0.00049487 & 0.00023359 & 0.00009798 & 0.00004746 \\ \hline

\end{tabular}%
}
\par\smallskip
\footnotesize
BS, MC, BO, and PPR denote the bootstrap SPR, Monte Carlo SPR,
Bonferroni SPR, and pointwise prediction region, respectively.
ECP, AV, and AVV denote the empirical simultaneous coverage probability,
average volume, and average variance of volume, respectively.
All procedures use the full PMSE $\bm G_1+\bm G_2$.
\end{sidewaystable}

\FloatBarrier

\end{document}